\documentclass[twocolumn,tighten]{aastex62}
\usepackage{amsmath}

\newcommand\be{\begin{equation}}
\newcommand\en{\end{equation}}

\begin{document}

\title{Global 3-D Simulations of Magnetospheric Accretion: II. Hot Spots, Equilibrium Torque, Episodic Wind, and Midplane Outflow }

\shorttitle{Magnetospheric Accretion: II}
\shortauthors{Zhu et al.}

\correspondingauthor{Zhaohuan Zhu}
\email{zhaohuan.zhu@unlv.edu}

\author[0000-0003-3616-6822]{Zhaohuan Zhu}
\affiliation{Department of Physics and Astronomy, University of Nevada, Las Vegas, 4505 S.~Maryland Pkwy, Las Vegas, NV, 89154, USA}
\affiliation{Nevada Center for Astrophysics, University of Nevada, Las Vegas, 4505 S.~Maryland Pkwy, Las Vegas, NV, 89154, USA}

\begin{abstract}
Global 3-D magnetohydrodynamical simulations have been conducted to study magnetospheric accretion around stars with various spin rates.  For slow rotators, characterized by a fastness parameter $\omega_s\lesssim 0.78$, the disk's inner edge at the magnetospheric truncation radius becomes unstable to the 
interchange instability, leading to intruding filaments which produce hot spots closer to the stellar equator.  Depending on spin rate, slow rotators can be in ``chaotic'' or ``ordered'' unstable regimes.
For fast rotators, the interchange instability is suppressed by the super-Keplerian rotation beyond the corotation radius, and hot spots are generated only through polar accretion.  Low- and mid-energy flux hot spots cover $\lesssim20\%$ and $\lesssim3\%$ of the surface, with faster rotators tending to produce hotter spots. Beyond the truncation radius, angular momentum transfers from the disk surface to the midplane, resulting in surface accretion and midplane outflow. The midplane outflow may transport thermally processed materials (e.g. those in chondrites) to the outer disk. Field inflation generates episodic winds with mass-loss rates $\sim$1\%--40\% of the accretion rate, depending on stellar spin.  Frequent magnetic reconnections lead to efficient star-disk coupling. We derive the torque exerted by the disk on the star as a function of stellar spin.  For fast rotators/propellers, both spin-down torque and disk wind rate increase dramatically with stellar spin. The equilibrium spin state occurs at $\omega_s\sim0.7$, with  wind/jet speeds ($\sim$500 km/s) and mass loss rates ($\sim10\%$ accretion rate) aligning with observations.  Most results are insensitive to disk thickness. Finally, we present testable predictions for how observables vary with stellar spin.            
\end{abstract}


\section{Introduction}
Magnetospheric accretion is the disk accretion onto a magnetized central object. 
This central object could be
neutron stars in Low Mass X-ray Binary systems (e.g. \citealt{Bildsten1997,Lewin2006}), young stars in Classical T-Tauri star systems (e.g. \citealt{Hartmann2016}), or even young planets (e.g. \citealt{Wagner2018,Haffert2019,Thanathibodee2019,Zhou2021}).  The strong magnetic fields of the central object halt
the radial inflow of the accretion disk, causing the disk material to climb and follow the magnetic field lines to reach the central object. This process controls the growth of the  central object, affects the evolution of the disk, produces the most energetic photons in the systems, and is potentially responsible for launching jets and outflows \citep{Fender2004}.

For T-Tauri stars, magnetospheric accretion plays important roles in star and planet formation. First, it produces strong X-ray and UV radiation from the accretion shocks  and hot spots. Close to the star, the disk material free falls to the stellar surface, guided by the stellar magnetic field lines. The collision between the infalling material and the stellar surface leads to strong shocks which can be as hot as 1 million Kelvin. The accretion shocks thus produce X-ray radiation that heats the underlying stellar photosphere to create ``hot spots'' at the stellar surface. The hot spots, around 7000 K, produce UV radiation \citep{CalvetGullbring1998,Hartmann2016}. By measuring the excess X-ray and UV radiation, or fluxes of certain atomic lines, we can determine the disk's accretion rates \citep{Hartmann1994, Muzerolle1998, Muzerolle2001}. Furthermore, since X-ray and UV photons are the main drivers for photochemistry, the accretion shocks and hot spots also affect molecules in the outer disks, which are being probed by JWST and ALMA. 

Due to the importance of UV radiation, the Hubble Space Telescope has spent 500 orbits obtaining UV spectra for 71 T-Tauri stars and brown dwarfs in the Hubble UV Legacy Library of Young Stars as Essential Standards (ULLYSES) program \citep{Roman-Duval2020}, the largest Director's Discretionary Program so far. Meanwhile, several ground-based programs have been conducted to obtain simultaneous optical spectra for these sources (e.g. ODYSSEUS \citealt{Espaillat2022}, PENELLOPE \citealt{Manara2021}). These programs start to reveal a wide range of physical conditions at the inner disk, from accretion to jet/wind \citep{Bouvier2023}. Particularly, hot spot properties are being constrained. With the more complete spectral coverage from UV to near-IR, hot spots with different temperatures have been discerned \citep{Espaillat2022,Pittman2022,Wendeborn2024}. It has been revealed that hot spots with low and medium energy fluxes (10$^{10}$ and 10$^{11}$ erg s$^{-1}$ cm$^{-2}$) dominate the total radiation energy budget, and the covering fraction for these hot spots is generally less than $20\%$.  By studying the time variabilities of UV and optical fluxes, \cite{Espaillat2021} further constrained the radial density gradient in a hot spot by using the movement of the hotspot into and out of view as it rotates along with the star. However, due to the complex processes involved in magnetospheric accretion,   MHD simulations are needed to understand connections between properties of hot spots, stars, and disks.

Magnetospheric accretion could affect the spin evolution of young stars (e.g. \citealt{Armitage1996}). Observations reveal that young stars exhibit a significant spread in stellar rotation periods, ranging from less than 1 day to approximately 10 days \citep{Herbst2002,Rebull2020,Serna2021}. Young stars with circumstellar disks generally rotate more slowly, suggesting that these disks affect stellar rotation. If the magnetic coupling between the star and the disk is efficient, the spin-up or spin-down torque is on the order of $T\sim\dot{M}(GM_* R_T)^{1/2}$, where $R_T$ is the magnetospheric truncation radius. Considering that the moment of inertial for a uniform sphere is $I=0.4M_*R_*^2$, the spin evolution timescale is $\sim I\Omega_*/T$. If $\Omega_*$ is around the orbital frequency at $R_T$, this timescale reduces to 0.4 $(M_*/\dot{M})(R_*/R_T)^2$. Thus, if $R_T\sim 3 R_*$, the spin evolution timescale is $\sim$ 10 times shorter than the accretion timescale, so that the presence of the disk can affect the stellar spin. However, for disks at later stages (e.g. $\dot{M}\lesssim10^{-8}M_\odot/yr$) or in cases of inefficient star-disk coupling \citep{Matt2005}, other mechanisms for stellar angular momentum loss (e.g. stellar wind) may be responsible for the observed rotation periods. Therefore, quantifying the magnetic coupling between the star and the disk is crucial for understanding the rotation of young stars. 

Magnetospheric accretion may also be responsible for the jets and winds in young stellar objects (YSOs). Jets in YSOs can reach speeds of 500 km/s \citep{Frank2014}, whose kinetic energy is equivalent to the gravitational potential energy at 2 solar radii around a solar-type star. If the launching mechanism roughly achieves energy equal-partition between various energy sources, the launching radius will be very close to the star, where magnetospheric accretion occurs. Thus, magnetospheric accretion simulation may also shed light on jet/wind launching

Finally, magnetospheric accretion influences the formation of close-in planets. The magnetospheric truncation radius can extend to 0.05 au. Furthermore, the stellar magnetic fields significantly affect the inner disk dynamics until the deadzone inner edge ($\sim 0.1-1$ au, depending on the stellar mass). Given that the exoplanet frequency within 0.1 au (10 days) is more than 0.2 and the frequency within 1 au is more than 1 \citep{ZhuDong2021}, the inner disk affected by magnetospheric accretion controls the formation and evolution of a large population of exoplanets. For example, dust evolution and gap opening in the inner disk can have direct impact on exoplanet formation \citep{Li2022,Li2024, Wu2019}. Overall, constraining the physical conditions (e.g. the disk surface density and turbulence level) at the inner disk during magnetospheric accretion is essential for studying the formation of close-in exoplanets.

The theoretical models of magnetospheric accretion are far from settled in many aspects. The seminal work by \cite{Ghosh1979II} assumes that the stellar fields can penetrate a large radial region of the disk, achieving a steady state when field dragging is balanced by dissipation. However, \cite{Aly1990, Lovelace1995, Uzdensky2002} point out that, with a more realistic field diffusion coefficient, stellar magnetic fields will be twisted by star-disk differential rotation. This twisting eventually leads to magnetic fields reconnection, with reconnected field loops being ejected from the system in a process known as ``field inflation'', potentially causing episodic accretion and wind. Meanwhile, \cite{Arons1986,Shu1994} suggest that the disk can be a good conductor, preventing magnetic field lines from penetrating through it, resulting in the field lines being pinched at the disk inner boundary. This model suggests a fan shaped wind from the disk inner boundary, called the ``X-wind''. 

Most previous simulations have adopted axisymmetric 2-D configurations \citep{Goodson1997,miller1997,Fendt2000,Zanni2013}, or used 3-D configurations with explicit viscosity to induce gas accretion \citep{Romanova2003,Romanova2004,Romanova2021}. These simulations have yielded important results, including interchange instability within the magnetosphere \citep{Kulkarni2008} and field inflating \citep{Goodson1997,miller1997, Romanova2009}.  Although interchange instability was first proposed and analyzed analytically \citep{Arons1976}, only the direct numerical simulations \citep{Kulkarni2008, Kulkarni2009} reveal that matter could penetrates very deep into the magnetosphere and even produce quasi-periodic unstable accretion. 
 However, previous simulations suffer some limitations. For example, 2-D ideal MHD simulations (e.g. \citealt{Romanova2018}) cannot capture the saturation of MRI and often leads to large scale disk variations due to axisymmetric MRI channel modes \citep{BalbusHawley1991,GoodmanXu1994}. 
The use of viscosity or/and resistivity in 2-D and 3-D MHD simulations  \citep{Ireland2021} introduces more free parameters. 

Recently, first-principle 3-D MHD simulations that resolve MRI turbulence have been carried out \citep{Romanova2012, Takasao2022,Zhu2024,Parfrey2023}. While earlier simulations focused on the magnetosphere, later simulations have reached a long enough timescale so that the disk outside the magnetosphere has also reached the quasi-steady state. While these first-principle MHD simulations have confirmed some results in previous viscous/resistive simulations, such as the interchange instability within the magnetosphere  \citep{Takasao2022,Zhu2024} and the field inflation \citep{Zhu2024, Parfrey2023}, they have also revealed new phenomena.  For example, outside the magnetosphere, a thick disk with a magnetized inflowing surface, similar to the disk threaded by net vertical magnetic fields \citep{ZhuStone2018,Takasao2018,Mishra2020M, Jacquemin2021}, has been observed.  Furthermore, large-scale magnetic bubbles, that orbit around the star at sub-Keplerian speeds, have been generated at the magnetospheric truncation radius \citep{Zhu2024, Parfrey2023}, similar to phenomena in magnetically arrested disks (MAD) around black-holes \citep{Porth2021}. Our previous work \citep{Zhu2024} only studied a disk around a non-rotating star. In this work, we extend it by running simulations with stars having a variety of stellar spin rates. We explore the impact of stellar rotation on hot spot distribution, interchange instability within the magnetosphere, magnetospheric truncation radius, disk accretion, jet/wind properties, and star-disk torque.
 
In \S2, we introduce key quantities for magnetospheric accretion. Our numerical model is described in \S 3. We present our results in \S 4. After discussions in \S5, we conclude the paper in \S 6.

\section{Angular momentum transport}
\label{sec:magstr}

Angular momentum transport equation is key to studying the stellar spin evolution, disk accretion, and wind:
\begin{equation}
\frac{\partial R\rho v_{\phi}}{\partial t}+\nabla_c\cdot\left(R\left(\rho{\bf v}v_\phi-{\bf B}B_{\phi}+P{\bf e_{\phi}}\right)\right)=0\,,\label{eq:ame}
\end{equation}
where $P$ is the total pressure, R is the horizontal distance to the rotational axis, and $\nabla_c$ is the divergence operator in cylindrical or spherical polar coordinates. 

If we choose a spherical region at the stellar surface with radius $r_s$, the stellar spin evolution can be derived using the torque ($T_{sd}$) between the star and the disk:
\begin{equation}
\frac{\partial \int_{0}^{r_s} R\rho v_{\phi} dV}{\partial t} = -\int_{r_s} R (\rho v_{r}v_{\phi}-B_r B_{\phi})dS\equiv -T_{sd} \,,  \label{eq:amestar}
\end{equation}
where the left-hand side represents the change in the angular momentum of the star, and $v_{\phi}(r) =\Omega_* r$ for a uniformly rotating star.
Considering that  angular momentum is changed by torque, we call $-R B_r B_{\phi}$ term as the magnetic torque and the $R \rho v_r v_{\phi}$ term as the hydro torque that includes both the turbulent torque ($R\rho v_r \delta v_{\phi}$)  and angular momentum carried by the mean flow ($R\rho v_r \langle v_{\phi}\rangle$ with $\langle v_{\phi}\rangle\equiv 1/(2\pi) \int_{0}^{2\pi}v_{\phi} d\phi$). If the star rotates slowly, disk accretion and star-disk interaction likely adds angular momentum to the star ($T_{sd}<0$) so that the star spins up. If the star rotates fast, angular momentum flows from the star to the disk ($T_{sd}>0$) so that the star spins down. We will measure $T_{sd}$ in our simulations and study its dependence on stellar rotation.

In the disk region that reaches a steady state, the second term in Equation \ref{eq:ame} becomes zero. In other words, if we define $S$ as an enclosed surface outside the star, we have
\begin{equation}
\int R\left(\rho{\bf v}v_\phi-{\bf B}B_{\phi}\right) \cdot d{\bf S} = const = T_{sd}\,, \label{eq:torqueconst}
\end{equation}
for any $S$ that reaches a steady state. Traditionally, $S$ is chosen as both sides of the disk surface (Figure 1 in \citealt{Ghosh1979III}), where the magnetic field lines  at the inner disk surface connect to the star and field lines at the outer disk surface open to infinity. The torque can thus be separated into the hydro torque from accretion and wind, the magnetic torque due to closed field lines, and the magnetocentrifugal wind torque from open field lines. However, in realistic MHD simulations, it is difficult to distinguish these different contributions due to the poorly defined disk surface. First, the disk is turbulent. Secondly, the disk is magnetically levitated and there is an extended surface region (up to $z\sim R$) that undergoes rapid surface accretion \citep{ZhuStone2018, Begelman2024}. For our simulations, it is easier to study the angular momentum budget by choosing $S$ as cylindrical surfaces around the central star. Then, we have:
\begin{equation}
    R \langle v_{\phi} \rangle \dot{M} +  \int R (\rho v_{R}\delta v_{\phi} - B_{R} B_{\phi}) dS = T_{sd}\,,
\end{equation}
at any $R$ where it is in a steady state. 
If we separate the vertical integration into the inflowing disk region ($|z|<z_d$) and the outflowing wind region ($|z|>z_d$), we have:
\begin{eqnarray}
    &R \langle v_{\phi} \rangle \dot{M}_{acc} +  \int_{-z_d}^{z_d} R (\rho v_{R}\delta v_{\phi} - B_{R} B_{\phi}) dS + \nonumber\\
     &R \langle v_{\phi} \rangle \dot{M}_{out} +  \int_{wind} R (\rho v_{R}\delta v_{\phi} - B_{R} B_{\phi}) dS = T_{sd}\,,
     \label{eq:torqueconst2}
\end{eqnarray}
so that the stellar spin-up/down torque ($T_{sd}$) is separated into accretion torque (including both mean flow and turbulent flow), disk magnetic torque, wind outflowing torque (mean flow+turbulent flow), and wind magnetic torque. 

Although this integral form is quite useful for studying stellar spin-up/down, it is rarely used for accretion disk studies, since $T_{sd}$ is normally unknown without properly modeling the stellar-disk interaction. Instead, the derivative form (Equation \ref{eq:ame}) is commonly used since the derivative of the constant $T_{sd}$ becomes 0.  With the definitions of $\delta v_{\phi}$ and $\langle v_{\phi}\rangle$ and the help of mass conservation equation, Equation \ref{eq:ame} can be rewritten as \citep{BalbusHawley1998}
\small
\begin{equation}
\frac{\partial R\rho \delta v_{\phi}}{\partial t}+\nabla_c\cdot\left(R\left(\rho{\bf v}\delta v_\phi-{\bf B}B_{\phi}+P{\bf e_{\phi}}\right)\right)+\nabla_p \left(R \langle v_\phi\rangle\right) \cdot\left(\rho{\bf v_p}\right)=0\,.
\label{eq:ame3}
\end{equation}
\normalsize
where $\nabla_p$ is the poloidal gradient in cylindrical or spherical coordinates.
In a steady state, Equation \ref{eq:ame3} suggests that the divergence of the turbulent and magnetic stress  leads to net mass flow. If $\langle v_\phi\rangle$ only depends on $R$, the equation can be written as
\small
\begin{align}
\frac{\dot{M}_{acc}}{2\pi R}\frac{\partial R \langle v_{\phi}\rangle}{\partial R}=&-\frac{1}{R}\frac{\partial}{\partial R}\left( R^2 \int_{z_{min}}^{z_{max}}\left(\langle \rho v_{R}\delta v_{\phi}\rangle-\langle B_{R}B_{\phi}\rangle\right)dz\right) \nonumber\\
&-R\left(\langle \rho v_{z} \delta v_{\phi}\rangle-\langle B_{z}B_{\phi}\rangle\right)\bigg |_{z_{min}}^{z_{max}}\label{eq:angcyl2}
\end{align}
\normalsize
where $\dot{M}_{acc}= 2\pi R\int \langle\rho v_{R}\rangle dz$ and we assume that the disk region extends from $z_{min}$ to $z_{max}$. 

Equation \ref{eq:angcyl2} is widely used in accretion disk studies (e.g., \citealt{Lesur2021}).  We refer to
$\langle \rho v_{R}\delta v_{\phi}\rangle$ and $-\langle B_{R}B_{\phi}\rangle$ as the radial Reynolds and Maxwell stress, and define the corresponding $\alpha$ parameters as
\begin{equation}
\alpha_{R}=\langle \rho v_{R}\delta v_{\phi}\rangle/\langle p\rangle\quad {\rm and}\quad \alpha_{M}=-\langle B_{R}B_{\phi}\rangle/\langle p\rangle \,.
\end{equation}
Stresses and the $\alpha$ parameters in spherical-polar coordinates (e.g. $\rho v_{r}\delta v_{\phi}$) can be defined in similar ways.
We can further define
the vertically integrated $\alpha$ parameter as
\begin{equation}
\alpha_{int}=\frac{\int T_{R\phi}dz}{\Sigma c_{s}^2}\,,\label{eq:alphaint}
\end{equation}
where $T_{R\phi}$ is the sum of both radial Reynolds and Maxwell stress.
Then, Equation \ref{eq:angcyl2} can be written as
\small
\begin{align}
&\dot{M}_{acc}=-\frac{2\pi}{\partial R \langle v_{\phi}\rangle/\partial R}\times\nonumber\\
&\left(\frac{\partial}{\partial R}\left( R^2 \alpha_{int}\Sigma c_{s}^2\right)+R^2\left(\langle \rho v_{z} \delta v_{\phi}\rangle-\langle B_{z}B_{\phi}\rangle\right)\bigg |_{z_{min}}^{z_{max}}\right)\,.\label{eq:mdot}
\end{align}
\normalsize
Equation \ref{eq:mdot} suggests that both the internal stress (or more specifically, the radial gradient of the $R$-$\phi$ stress) and the surface stress can lead to accretion. We will
measure these stresses and $\alpha$ values directly from our simulations.

\section{Method}

With a simplified thermodynamics (e.g. locally isothermal equation of state with a known radial temperature slope) and stellar properties (e.g. spinning axis aligning with the disk rotation axis), there are only 3 dimensionless parameters to specify a steady magnetospheric accreting system: the ratio between the magnetospheric truncation radius and the stellar radius, the ratio between the truncation radius and the corotation radius, and the disk aspect ratio at the truncation radius. If we ignore the feedback from the accreting star to the disk,  the last two parameters are key parameters which will be explored in this work.

We solve the  ideal magnetohydrodynamic (MHD) equations  using Athena++ \citep{Stone2020}. Our simulation setup follows \cite{Zhu2024} and is summarized below. We adopt a Cartesian coordinate system ($x$, $y$, $z$) with mesh-refinement  to include both the central magnetized star and the accretion disk. This setup allows us to study funnel flows at the pole and the accretion disk at the equator equally well. Our simulation domain extends from -64$R_0$ to 64$R_0$ in each $x$, $y$, and $z$ direction, with 320 grid cells in each direction at the root level. $R_0$ is the code length unit\footnote{With our choice of field strength, $R_0$ turns out to be approximately twice the magnetospheric truncation radius ($R_T$), which is different from \cite{Zhu2024} where a stronger stellar field results in $R_0\sim R_T$.}, and the orbital frequency at $R_0$ is $\Omega_0$. Thus, the time unit is $1/\Omega_0$ and the orbital period at $R_0$ is $T_0=2\pi/\Omega_0$.  The stellar radius, denoted as $r_{s}$, is chosen as 0.1 $R_0$. 
Static mesh-refinement has been adopted with the fourth level (cells with 2$^4$ times shorter length) at [-8, 8]$R_0\times$[-8, 8]$R_0\times$[-1, 1]$R_0$ for the $x\times y\times z$ domain. 
Within this fourth level domain, one additional higher level is used for every factor of 2 smaller domain until the seventh level at [-$R_0$, $R_0$]$\times$[-$R_0$, $R_0$]$\times$[-0.125$R_0$, 0.125$R_0$]. For a disk having the aspect ratio of 0.1, the disk scale height is resolved by 16 to 32 grid cells at the disk region between $R=0.5 R_0$ to $R=8 R_0$. At the finest level, the cell size is 0.003125 $R_0$. 

A hydrostatic disk is initially set up with the midplane density  at $R_0$ as $\rho_0$=1 in code units. The disk surface density initially follows $\Sigma\propto R^{-1}$. The temperature is assumed to be constant on cylinders and follows $T\propto R^{-1/2}$.  Although an adiabatic equation of state with $\gamma$=1.4 has been adopted, the temperature is reset to the initial temperature during each timestep so that this is equivalent to adopting a locally isothermal equation of state \citep{Zhu2015c,Miranda2020}.  For most of our simulations, the disk aspect ratio ($h$) is chosen as 0.1 at $R_0$ so that the most unstable MRI mode in the disk is sufficiently resolved \citep{ZhuStone2018}. Since a realistic protoplanetary disk has a scale height of $\sim$ 0.04 at the simulated region, we have carried out one additional simulation with the aspect ratio of 0.05 at $R_0$, which will be presented in \S \ref{sec:thindisk}.

The setup of the star is described in detail in \cite{Zhu2024}, with slight modifications. The density of the star is set to $10^5$ $\rho_0$. The star has a uniform rotation frequency of $\Omega_{s}$. Both the density and velocity are fixed to the initial values within $r=0.16 R_0$, which includes half a million grid cells. The star also has initial dipole  magnetic fields ${\bf B}=\nabla\times{\bf A}$ with
\begin{equation}
{\bf A}=\frac{{\bf \overline{m}}\times {\bf r}}{r_c^3}\,,
\end{equation}
where $r_c=max(r, r_{s})$ to avoid the singularity at r=0.
Thus, the magnetic fields beyond the star are:
\begin{equation}
{\bf B}({\bf r})=\frac{3{\bf r}({\bf \overline{m}}\cdot {\bf r})}{r^5}-\frac{{\bf \overline{m}}}{r^3}\,. \label{eq:Bdipole}
\end{equation}
Note that the vacuum permeability constant is assumed to be 1
in Athena++, and thus the magnetic pressure is simply $B^2/2$ in code units. We choose ${\bf {\overline m}}$=-0.0089${\bf e_z}$ so that the plasma $\beta=2P/B^2$ at $R=R_0$ is 250 initially.  Since the star is fixed to be rotating at a constant $\Omega_s$ within 0.16$R_0$, the ${\bf v}\times{\bf B}$ term in the induction equation drives the rotation of the stellar magnetic fields.  Since the plasma $\beta$ is less than one at 0.16 $R_0$, the rotating magnetic dipole leads to the rotation of the magnetosphere. 

We employ a density floor that varies with position: 
\begin{equation}
\rho_{fl}=\rho_{fl,0}\left(\frac{r}{R_0}\right)^{-2.25}+\rho_{flm,0}\left(\frac{r}{R_0}\right)^{-5.5} \,,\label{eq:floor}
\end{equation}
where $\rho_{fl,0}=10^{-6}\rho_0$, and $\rho_{flm,0}$=1.6$\times 10^{-6}\rho_0$. 
When $\rho_{fl}$ becomes smaller than $10^{-9}\rho_0$, we choose $10^{-9}\rho_0$ as the density floor. To transition from the disk to the star, some smoothing profiles for the density and velocity have been adopted as in \cite{Zhu2024}.

 The initial velocity is the same as that in \cite{Zhu2024}. The velocity in the whole domain follows that of a rotating disk, except for a small spherical region ($r<0.5 R_0$) around the star (Appendix). We have also run a simulation with the initial velocity setup following \cite{Romanova2002}, and obtained almost identical results. 

For data analysis, we transform all quantities into spherical-polar or cylindrical coordinates after the simulations are completed.  We use ($R$, $\phi$, $z$) to denote positions in cylindrical coordinates and
($r$, $\theta$, $\phi$) to denote positions in spherical-polar coordinates.  In both coordinate systems, 
$\phi$ represents the azimuthal direction, and the direction of disk rotation has positive $v_{\phi}$. 

\section{Results}

\begin{figure*}[t!]
\includegraphics[trim=0mm 5mm 0mm 0mm, clip, width=7.in]{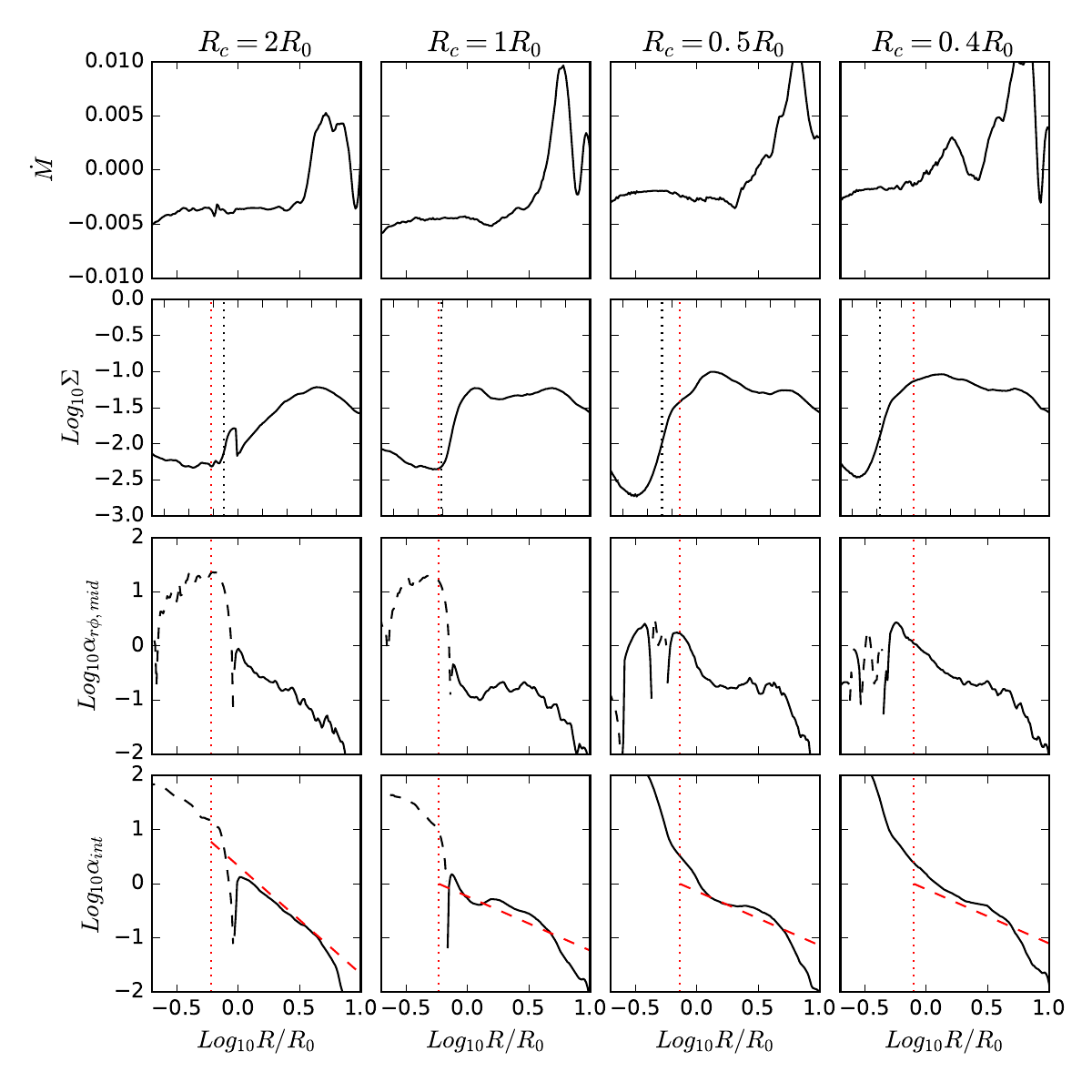}
\figcaption{ The vertically integrated disk mass accretion rate, surface density, midplane $\alpha$, and vertically integrated $\alpha$ for disks with different stellar spin rates at the end of the simulations.  Dashed black curves represent negative values. All quantities are time-averaged over the last 3 T$_0$ using 30 snapshots. The vertical red lines indicate  $R_{T,ana}$. The vertical black lines in the $\Sigma$ panel indicate $R_T$. The red 
dashed line in the leftmost bottom panel represents $\alpha_{int}=6 (R/R_{T,ana})^{-2}$, while the red dashed lines in the right three panels represent $\alpha_{int}=(R/R_{T,ana})^{-1}$.  
\label{fig:oned}}
\end{figure*}

We have carried out four simulations with different stellar spin rates: $\Omega_s$=0.354, 1, 2.8, 3.95 $\Omega_0$, corresponding to corotation radii $R_c$= 2, 1, 0.5, 0.4 $R_0$, respectively. These four simulations are denoted as Rc2, Rc1, Rc0p5, and Rc0p4.  A dimensionless parameter, called fastness parameter $\omega_s\equiv \Omega_s/\Omega_{K}(R_T)=(R_T/R_c)^{3/2}$, is often used to describe the relative importance between stellar spin and disk accretion. Since $\omega_s$ depends on the magnetospheric truncation radius $R_T$, it has to be derived from the simulation (Table \ref{tab:summary}) and changes with time. Using the measured truncation radius at the end of the simulations ($R_{T}$=0.76, 0.61, 0.39, 0.42 $R_0$, the formal definition of $R_T$ is in \S 4.4), $\omega_s=$0.24, 0.48, 1.05, and 1.08, respectively. 
To introduce perturbations to the axisymmetric disk initially, we include a planetary perturber in the disk for 17 $T_0$. We then remove the planetary perturber and continue the simulation until 34, 30.4, 32.8, and 31.3 $T_0$, respectively. This is equivalent to $\sim$ 500 Keplerian orbits at the fixed stellar region of 0.16 $R_0$. We have also carried out a $R_c$=0.5 $R_0$ simulation without any initial planetary perturber, finding that initial perturbations have little impact on the results. 

Most of our simulations have reached steady states within $R\sim 3 R_0$, as shown in the top panels of Figure \ref{fig:oned}, except for the fast rotator case (Rc0p4), where the accretion rate is affected by a strong and highly variable wind.  The estimated accretion rates are $\sim$ -0.004, -0.005, -0.0025, -0.002 in the code unit of $\rho_0 R_0^3/T_0$, respectively.  With a given disk accretion rate and stellar dipole fields, the magnetospheric radius can be estimated analytically as
\begin{equation}
R_{T,ana}=r_{*}\left(\frac{B_*^4 r_*^5}{2GM_*\dot{M}^2}\right)^{1/7}\, \label{eq:RT}
\end{equation}
in C.G.S. unit, where $B_*$ is the dipole magnetic field strength at the stellar surface ($r_*$) on the equator. 
With our stellar dipole strength (Equation \ref{eq:Bdipole}), Equation \ref{eq:RT} reduces to
\begin{equation}
R_{T,ana}=\left(\frac{\overline{m}^4 \left(4\pi\right)^2}{2GM_*\dot{M}^2}\right)^{1/7} \,.\label{eq:RTcode}
\end{equation}
With the measured accretion rates, we have $R_{T,ana}$=0.61, 0.57, 0.70, 0.74 $R_0$, respectively. 
$R_{T,ana}$ are plotted as vertical dotted lines in Figure \ref{fig:oned}. In the  $\Sigma$ panels, we also plot $R_{T}$ which traces the disk inner edge very well.
For fast rotators, $R_{T}$ is significantly smaller than $R_{T,ana}$ and close to $R_c$ (more discussion in \S 4.4). For the slowest rotator, the surface density increases with radius, which is led by the fast decreasing of $\alpha$ with radius. \cite{Zhu2024} showed that for the non-rotator, $\alpha_{mid}\propto R^{-1.5}$ and  $\alpha_{int}\propto R^{-2}$ outside the truncation radius. We find that such fitting works well for our slowest rotator too ($\alpha_{int}=6(R/R_{T,ana})^{-2}$). Thus, to maintain a steady state, $\Sigma$ needs to follow $\Sigma\propto R$, as shown in the figure.
For faster rotators, the $\alpha$ curves are slightly flatter, and $\alpha_{int}=(R/R_{T,ana})^{-1}$ provides a good fit. Overall, this large $\alpha$ is consistent with that in  \cite{Zhu2024}, implying a very low disk surface density and slow planet migration at the inner MRI active disk \citep{Zhu2024}. 

Within the magnetosphere, the slow rotors have negative $\alpha_{int}$ values. This occurs because the disk outside $R_T$ rotates faster than the slowly rotating star, stretching the stellar magnetic field lines forward along the disk's rotation direction. This field orientation is opposite to the magnetic field lines from the Keplerian differential rotation. Overall, within the magnetosphere, angular momentum is transferred inwards from the disk to the star, spinning up the star. For fast rotators, $\alpha_{int}$ is positive, so that the star loses angular momentum to the disk and spins down. 

In the following subsections, we  will study magnetospheric accretion in greater details, including the filaments and outflow at the midplane (\S 4.1), hot spots on the surface of the star (\S 4.2), wind from the disk surface (\S 4.3), the magnetospheric truncation radius (\S 4.4), and the stellar spin-up/down torque (\S 4.5).

\subsection{From Intruding Filaments to Outflowing Midplane}
\begin{figure*}[t!]
\includegraphics[trim=0mm 0mm 0mm 0mm, clip, width=7.in]{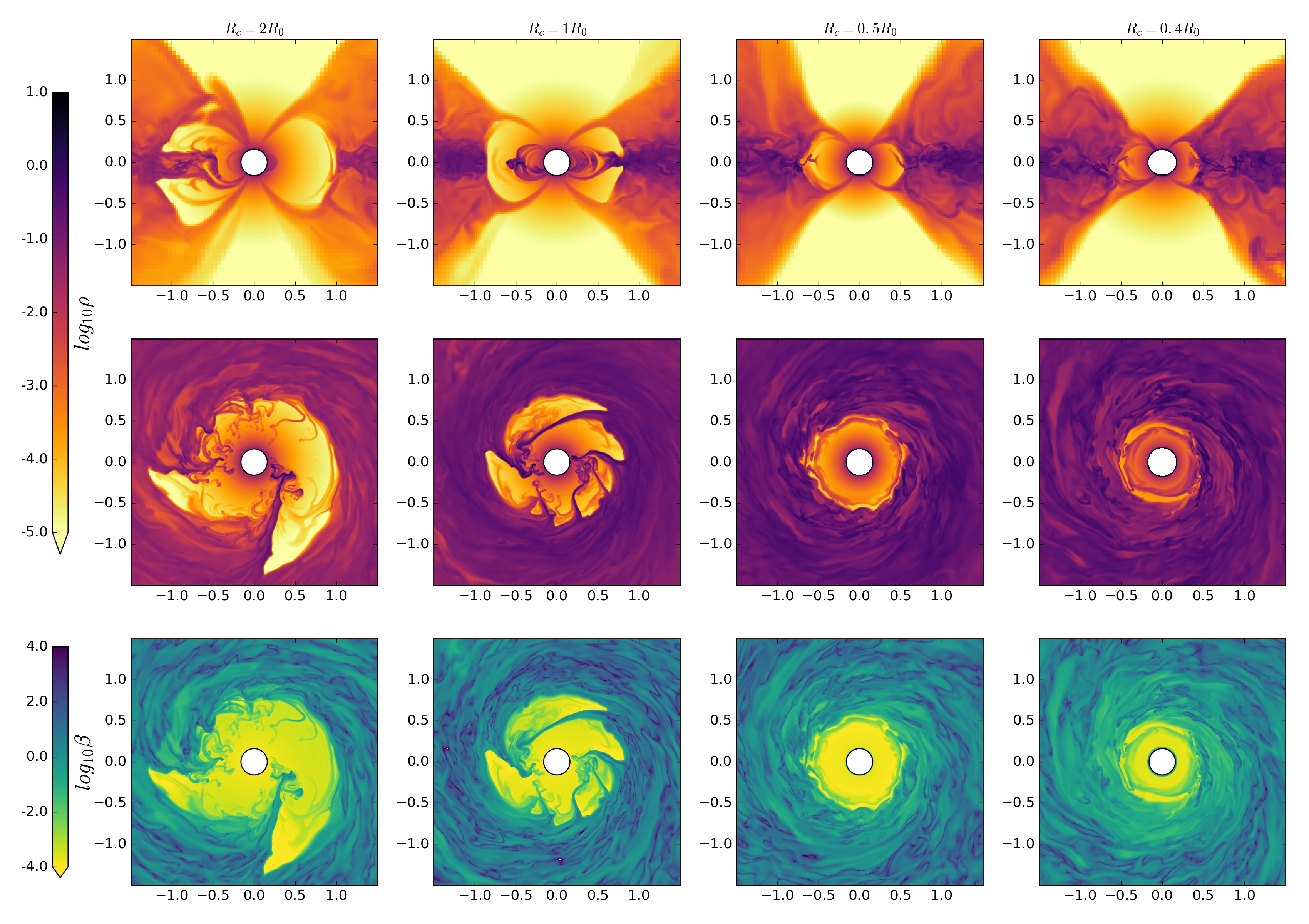}
\figcaption{ 
Poloidal and midplane cuts of density and plasma $\beta$ at the end of the simulation for different cases. The movie can be found  at \url{https://doi.org/10.6084/m9.figshare.26948260.v2}.
  \label{fig:mid2d}}
  \end{figure*}
  
\begin{figure*}[t!]
\includegraphics[trim=0mm 9mm 0mm 8mm, clip, width=7.in]{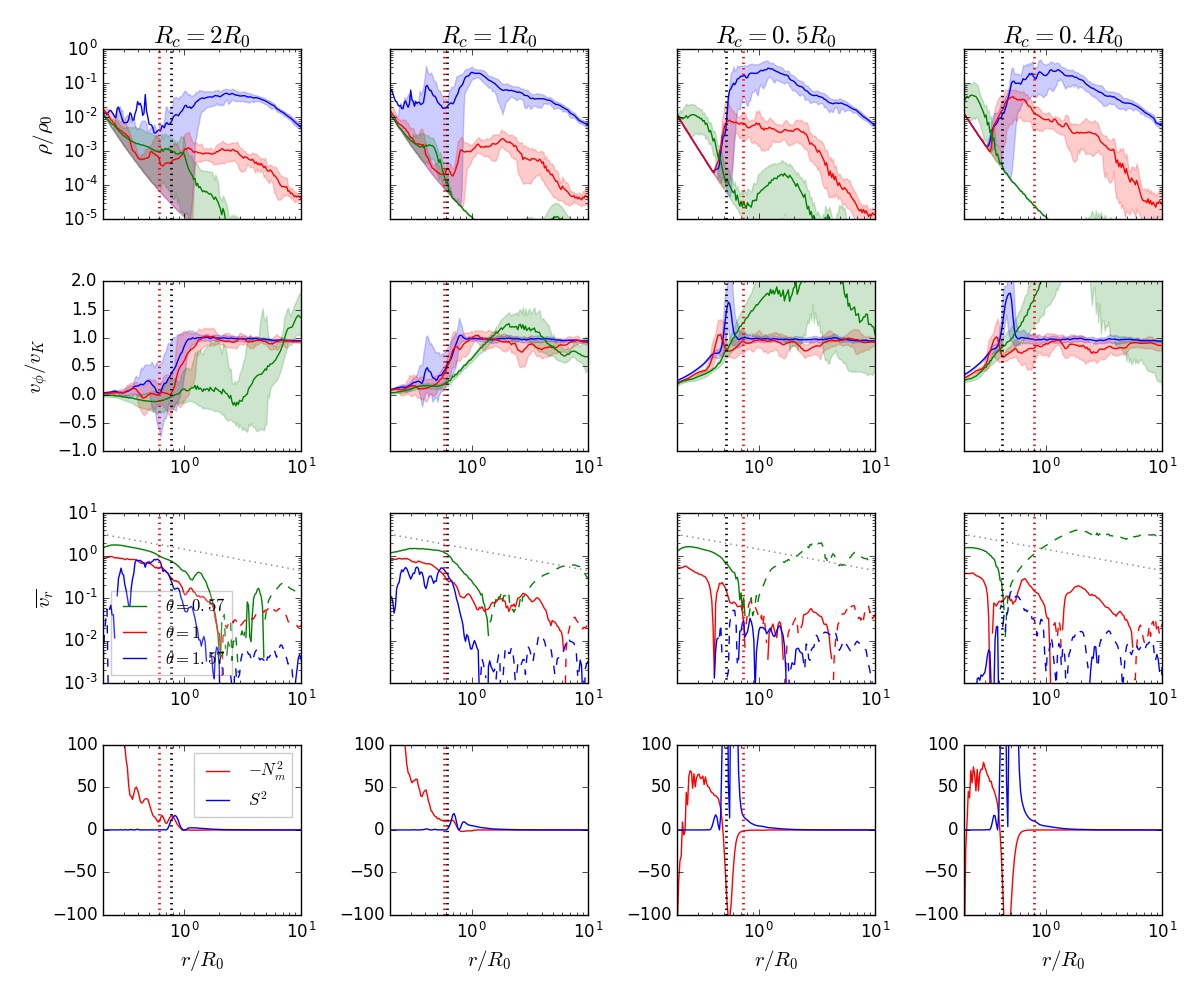}
\figcaption{ 
The azimuthally averaged density,  azimuthal velocity, and radial velocity (top three rows) along different $\theta$ directions (blue: $\theta=1.57$, red: $\theta=1$, green: $\theta=0.57$) at the end of the simulations for different cases (left to right panels). The shaded areas in the top two rows represent the range between the 10th and 90th percentiles of all data along the azimuthal direction. The vertical black ($R_{T}$) and red ($R_{T,ana}$) dotted lines indicate the truncation radius directly measured in simulations and derived from accretion rates, respectively. The dotted tilted lines in the $v_r$ row represent the free-fall speed. The bottom row shows the ``interchange instability'' growth term $-N_{m}^2$ and damping term $S^2$ along the midplane. 
\label{fig:mid1d}}
\end{figure*}  
  
A faster stellar spin suppresses the development of intricate features around the truncation radius, as shown in Figure \ref{fig:mid2d}. Both the intruding filaments and the magnetic bubbles outside are suppressed for fast rotators. 

For a slow rotator as in \cite{Zhu2024}, the region  around $R_T$ rotates at a speed smaller than the Keplerian speed (the left panel in the middle row of Figure \ref{fig:mid1d}),  so that, in the radial direction, the region inside $R_T$ will be magnetically supported instead of rotationally supported. Magnetically supported region is subject to ``interchange instability'' \citep{Kruskal1954,Newcomb1961,Arons1976}, a type of Rayleigh-Taylor instability in magnetized fluid.   Such instability has recently been confirmed in laboratory experiment \citep{Burdonov2022}. 
The instability leads to penetrating ``fingers'' into the magnetosphere (middle left panels in Figure \ref{fig:mid2d} and \citealt{Kulkarni2008}), resulting in a modest density at the midplane (upper left panel in Figure \ref{fig:mid1d}). 
The intruding filaments are continuously lifted out of the midplane as they move inward (upper left panels in Figure \ref{fig:mid2d}). If the star is sufficiently small compared to the size of the magnetosphere, eventually all filaments will be lifted up and no filaments could reach the star at the equator, leaving the midplane devoid of material. Otherwise, some filaments will hit the star directly at the equator. At the midplane outside $R_T$, large-scale magnetic bubbles occasionally appear \citep{Zhu2024,Parfrey2023}, which is an extreme manifestation of the interchange instability.  These bubbles orbit around the star at sub-Keplerian speeds and are highly magnetized (magnetically buoyant). As the bubbles move to outer radii, they are sheared away, similar to those in magnetically arrested disks (MAD, \citealt{Tchekhovskoy2011,Porth2021}).

When the stellar spin increases (towards the right panels in Figure \ref{fig:mid2d}),   the region around $R_T$ becomes more rotationally supported, and ``interchange instability'' is suppressed \citep{Spruit1995, Romanova2008, Blinova2016, Takasao2022}. Consequently, the midplane within $R_T$ is devoid of mass (its density follows the density floor in the simulation), and there is a sharp density drop at $R_T$ (upper right panels in Figure \ref{fig:mid1d}). The suppression of the instability could also be intuitively understood based on the azimuthal velocity profile. For fast rotators where $R_c$ is close to $R_T$, the material outside $R_c$ is strongly propelled by the stellar fields, and it can be accelerated to super-Keplerian speed.  This super-Keplerian rotation is clearly shown in the right $v_{\phi}$ panel of Figure \ref{fig:mid1d}. Any material leaking into $R_T$ at the midplane will be centrifugally repelled out, preventing the midplane intruding. This midplane outflow at $R_T$ for fast rotators is evidenced by the blue curve in the rightmost $v_r$ panel of Figure \ref{fig:mid1d}.

Quantitatively, the general ``interchange instability'' criterion that considers the fluid's shear is given in \cite{Spruit1995}. The growth term ($-N_m^2$) is related to the magnetic buoyancy frequency $N_m$ and the damping term ($S^2$) is related to the shear rate $R\frac{d\Omega}{dR}$:
\begin{eqnarray}
-N_m^2= g_m \frac{d}{dR}ln\frac{\Sigma}{B_z}\,,\label{eq:Nm}\\
S^2= 2 \left(r\frac{d\Omega}{dR}\right)^2\,.
\end{eqnarray}
When $-N_m^2>S^2$, the instability grows super-linearly.  We plot these two terms at the disk midplane in the bottom row of Figure \ref{fig:mid1d}.  $g_m$ is the magnetic acceleration at the midplane which can be calculated using $g_m=(\Omega_K^2-\Omega^2)R$, $\Sigma$ is the vertically integrated surface density, and $B_z$ is the midplane vertical magnetic field strength. $\Sigma$, $v_{\phi}$, and $B_z$ are averaged over 30 snapshots for the last 3 orbits at $R_0$. When $S=0$, the fluid is unstable if it is partially magnetically supported ($g_m>0$) and $\Sigma/B_z$ increases with R. The left two panels of Figure \ref{fig:mid1d} show that, for slow rotators with $R_c>R_{T,ana}$, the region around the truncation radius has positive $-N_m^2$ with positive $g_m$ (magnetically supported). Outside the truncation radius, the shear could stabilize the flow. However, at the truncation radius (either $R_{T,ana}$ or $R_{T}$), the growth term is higher than the damping term, which leads to filaments within the cavity. For fast rotators with $R_c<R_{T,ana}$ (the right two panels), the large $v_{\phi}$ around $R_T$ leads to large shear
and high  $S^2$ values. At the same time, $-N_m^2$ becomes negative since $g_m$ is negative with the super-Keplerian motion.  Overall, for fast rotators, the super-keplerian rotation around $R_T$ makes this region stable to ``interchange instability''. We will examine this condition in more detail for the equilibrium spin state in \S 5.1.

\begin{figure*}[t!]
\includegraphics[trim=0mm 98mm 0mm 2mm, clip, width=7.in]{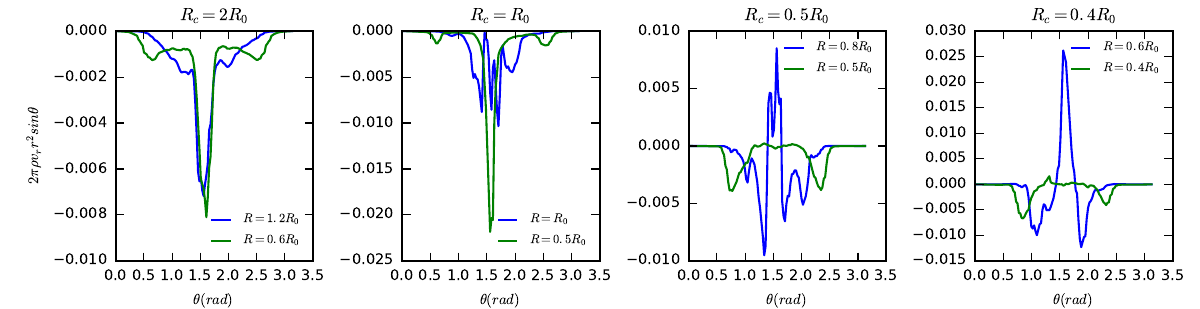}
\figcaption{ Radial mass flux along the $\theta$ direction (in radians) at the end of the simulation for different cases. All quantities are time-averaged over the last 3 T$_0$ using 30 snapshots. 
\label{fig:mdottheta}}
\end{figure*}

Stellar spin also affects the flow structure outside the magnetospheric truncation radius. For the slowest rotator with $R_c=2 R_0$, disk material at all heights flows inwards within $2 R_0$ (the left $v_r$ panel of Figure \ref{fig:mid1d}). Most mass flux occurs around the midplane at both $R_T$ ($\sim$ 0.6$R_0$) and 2$R_T$  (the leftmost panel in Figure \ref{fig:mdottheta}). At $R_T$, the midplane accretion is primarily due to the penetrating filaments. At 2$R_T$ ($\sim$ 1.2$R_0$),  which is still inside the corotation radius $R_c$ (2 $R_0$), the disk rotates faster than the star. Thus, the magnetic field lines connection the star and the disk transport angular momentum from the disk to the star, causing the star to spin up and the disk to accrete inwards. 
For the rotator with $R_c=R_0$ (the second panel in Figure \ref{fig:mdottheta}), the midplane still accretes inwards at the truncation radius ($0.5 R_0$), with less inward accretion at the corotation radius $R_0$. For fast rotators (the rightmost two panels in Figure \ref{fig:mdottheta}), accretion generally occurs at high latitudes. Accretion at the midplane is negligible at $R_T$, and the midplane flows outwards outside the truncation radius (also in the right $v_r$ panels in Figure \ref{fig:mid1d}). Such outward midplane flow is also observed when the disk and the star are in the equilibrium spin state (\S 5.1).  This is dramatically different from the expectation that outflow occurs at the surface and accretion occurs at the midplane. In our simulations with fast rotators, the disk has three layers: the outflowing wind region, the surface accretion region, and the midplane outflow region. The surface accretion seems to be a unique feature of 3-D MHD simulations, different from previous 2-D $\alpha$-disk simulations. Although the midplane outflow has also been observed in previous 3-D MHD simulations with net vertical magnetic fields \citep{SuzukiInutsuka2014,ZhuStone2018},  the midplane outflow in magnetospheric accretion simulations is affected, at least indirectly, by the stellar spin. We will examine the accretion structure in more detail at \S 4.5.

\subsection{Hot Spots}

\begin{figure*}[t!]
\centering
\includegraphics[trim=0mm 20mm 0mm 17mm, clip, width=5.in]{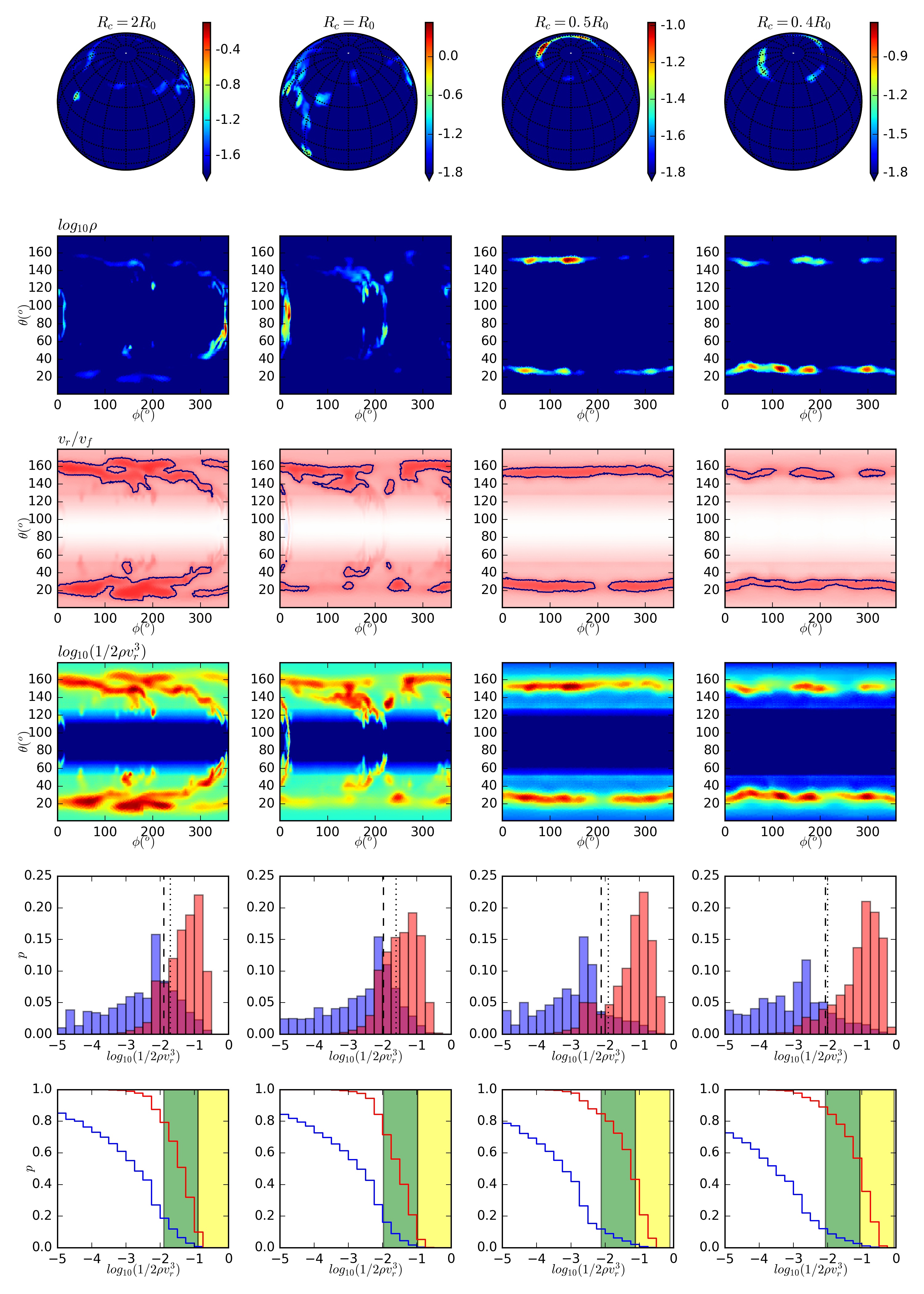}
\figcaption{ Various quantities at $r=0.2 R_0$ for different cases (left to right panels). The top two rows display $log_{10}\rho$ mapped to a sphere (top row) and in the $\phi$-$\theta$ plane (second row).  The third row shows the radial velocity normalized to the free-fall velocity, with colors ranging from red (-1) to white (0); the blue contour indicates a value of 0.5.   The fourth row shows the radial kinetic energy flux. The fifth row presents the area covering fraction of hot spots with different radial kinetic energy fluxes (blue bars) and the contribution of each energy flux bin to the total energy flux (red bars).  The vertical dotted lines represent the energy fluxes derived as 0.5 $GM_*\dot{M}/(4\pi r^3)$ using the measured accretion rates (0.004, 0.005, 0.0025, 0.002), while the dashed lines show the mean energy fluxes measured from the simulations at $r=0.2 R_0$. The bottom row shows the cumulative distribution of the radial kinetic energy fluxes from the fifth row. The red curves show the fraction of total energy released above certain energy flux bin, while the blue curves show the covering fraction of hot spots above that energy flux bin. The green shaded regions represent the low energy flux bins, ranging from  the mean energy flux (dashed line in the fourth row) to 10 times higher. The yellow shaded regions mark mid-energy flux bins, which is 10 times higher than the low-energy flux bins.
\label{fig:sphere}}
\end{figure*}

The intruding filaments affect hot spots on the surface of the star. The temperature of the hot spot is determined by the energy flux of the colliding filament:
\begin{equation}
T_{spot}=\left(\frac{3\rho v^3}{8\sigma}\right)^{1/4}\,,\label{eq:hotspot}
\end{equation}
where $\rho$ and $v$ are the density and velocity of the infalling filament \citep{Hartmann2016}. Figure \ref{fig:sphere} shows the distributions of
the density, velocity, and radial kinetic energy flux of the infalling filament at r=0.2 $R_0$.  Here, we assume that the stellar surface is at r=0.2 $R_0$, roughly half of the magnetospheric truncation radius ($R_T\sim$0.4-0.5 $R_0$). Slow rotators have high density regions close to the equator due to the penetrating filaments from the interchange instability. However, the radial infall speed is low at the equator since the radial motion at the equator is perpendicular to the direction of the magnetic fields. Material has to ``squeeze'' its way between magnetic field lines through interchange instability. On the other hand, the infalling material close to the pole follows along the magnetic field lines, allowing it to reach very high speed ($>0.5$ free-fall speed). Since the hot spot temperature (Equation \ref{eq:hotspot}) is more sensitive to velocity ($v^{3/4}$) than density ($v^{1/4}$), most energetic hot spots are around the pole for all the cases, as shown in the radial energy flux panels. On the other hand, slow rotators' hot spots extend closer to the equator compared with fast rotators.

The covering fractions of hot spots are shown in the bottom two rows of Figure \ref{fig:sphere}.  The blue bars in the fifth row show the area covering fraction of hot spots within different radial kinetic energy flux bins, while  the red bars show the contribution of each energy flux bin to the integrated total energy flux. The bins are divided uniformly in the logarithmic space. Summing the values of all bars with the same color equals 1. The vertical dotted line represents the energy flux derived using 0.5 $GM_*\dot{M}/(4\pi (0.2 R_0)^3)$ with the measured accretion rate: $\dot{M}=$0.005, 0.005, 0.0025, 0.0025 respective. It represents the mean energy flux if we assume that half of the accreting material's gravitational energy is released uniformly at the stellar surface.
By comparison, the dashed line represents the mean energy flux directly measured from the simulation by averaging $\rho v^3$ throughout the surface. The kinetic energy of the infalling material will be converted to the thermal energy of the hot spots (Equation \ref{eq:hotspot}) and eventually radiates away. Thus, the distribution of this infalling energy flux also represents the distribution of hot spots' radiation energy. The red bars show that most of the radiation comes from hot spots with an energy flux extending from just above the mean energy flux up to
1-2 orders of magnitude higher. The faster the star spins, the more radiation comes from hotter spots on the surface (having higher energy fluxes). 

The bottom row of Figure \ref{fig:sphere} shows the cumulative distribution of the kinetic energy fluxes, derived by integrating both red and blue curves in the fifth row starting from the highest energy flux. Thus, the red curve shows the fraction of total energy released above a certain energy flux bin, while the blue curve shows the covering fraction of hot spots above  a certain energy flux bin. The green shaded region labels the low-energy flux hot spots, defined as having an energy flux from the mean energy flux (dashed line in the fourth row) to 10 times higher. The yellow shaded region represents the mid-energy flux hot spots, which have 10 times higher energy flux than the low-energy flux hot spots.  The low- and mid-energy flux hot spots dominate most energy output. The covering factor of low-energy flux hot spots is $\lesssim$20\% and that of mid-energy flux hot spots is $\lesssim$3\%. For slow rotators, the low-energy flux hot spots dominate. For fast rotators, there is more contribution from mid-energy flux hot spots, e.g., its $\sim 3\%$ covering fraction and $\sim 60\%$ energy fraction for the Rc0p4 case.

\begin{figure*}[t!]
\includegraphics[trim=0mm 15mm 0mm 10mm, clip, width=7.in]{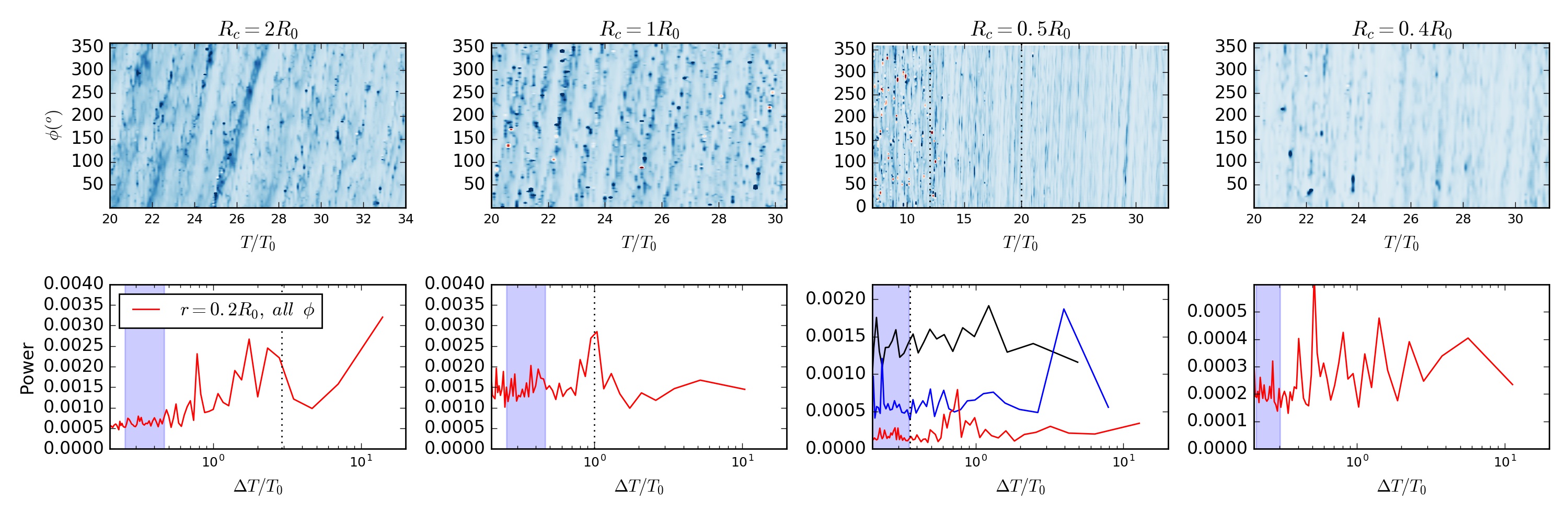}
\figcaption{ Top panels: Space-time diagrams of $\rho v_r$ at various $\phi$ directions at $r=0.2 R_0$ for all four cases. Bottom panels: Averaged periodograms of the mass accretion rates with time at $r=0.2 R_0$ across 256 different $\phi$ directions from  0 to 2$\pi$.   The mass accretion rate with time for each $\phi$ direction corresponds to a horizontal  cut in the top panels. For the Rc0p5 case, periodograms are calculated for three different timespans: black curve $[7 T_0, 12 T_0]$, blue curve $[12 T_0, 20 T_0]$, and red curve $[20 T_0, 32.8 T_0]$).
The dotted lines in the bottom panels indicate the stellar rotation periods, while the blue shaded regions represent the orbital periods at the magnetospheric truncation radius throughout the simulation.  
\label{fig:spacetime}}
\end{figure*}

To study the rotation of the hot spots around the stellar spin axis, we integrate $\rho v_{r}$ along the poloidal direction ($\theta$ direction) to derive the mass flux at different azimuthal angles ($\phi$). We then plot this mass flux at different $\phi$ angles against time, as shown in the upper panel of Figure \ref{fig:spacetime}. When a dominant hot spot rotates around the axis, its mass flux will leave a tilted stripe on this space-time diagram. The orbital frequency of the hot spots can be derived from the slope of the tilted stripe. Evidence of rotating hot spots is apparent in the figure. To determine the rotation period, we calculate the periodogram for every $\phi$ angle and then average the periodograms to reduce noise (bottom panels of Figure \ref{fig:spacetime}).  The vertical dotted lines in the bottom panels represent  the stellar spin periods. 
For slow rotators (the left two panels), the periodograms have peaks at the stellar spin frequency, indicating that some hot spots co-rotate with the star.  However, there are many peaks that do not correspond to the spin period. For our slowest rotator (the leftmost panel), the peaks at shorter periods are even stronger than the peak at the spin period. This is likely due to the fewer but prominent filaments and the magnetic bubbles. This resembles the ``ordered unstable regime'' in \cite{Blinova2016}. When the stellar spin increases, the peak at the stellar period dominates (the Rc1 case), resembling the ``chaotic unstable regime'' in \cite{Blinova2016}.
For cases with even faster spins, the accretion is in the stable regime.  More quantitative comparison with \cite{Blinova2016} will be given in \S 5.4. For fast rotators, the additional peaks likely correspond to episodic winds from field inflation, and they are typically at several times the orbital period at the truncation radius ($T(R=0.5R_0)=0.35 T_0$). Overall, caution is needed when using peaks in the periodogram to determine the stellar spin period or to detect the close-in planets in the disk. 

\subsection{Wind}

\begin{figure*}[t!]
\centering
\includegraphics[trim=0mm 9mm 0mm 0mm, clip, width=4.8in]{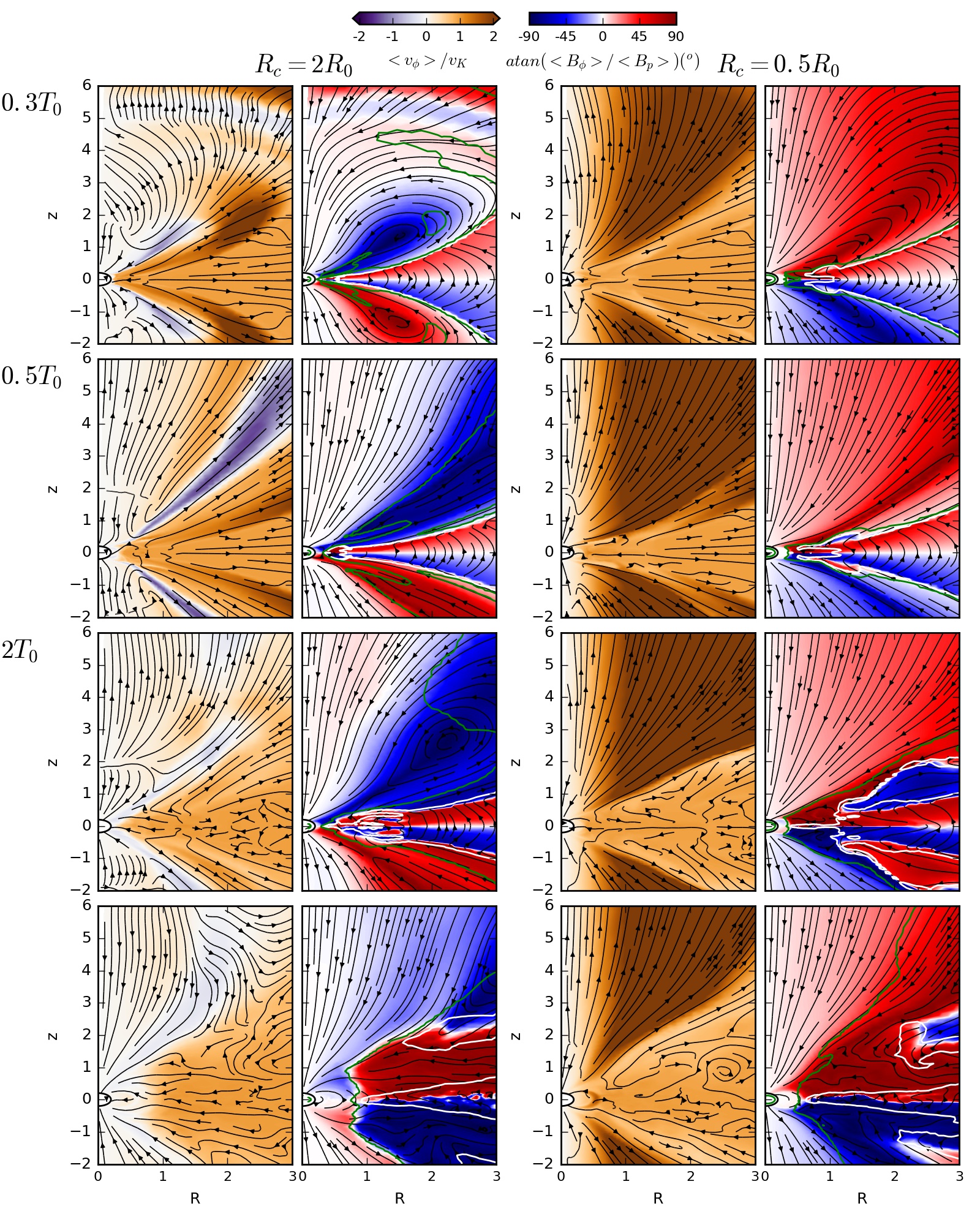}\\
\includegraphics[trim=0mm 72mm 0mm 53mm, clip, width=4.8in]{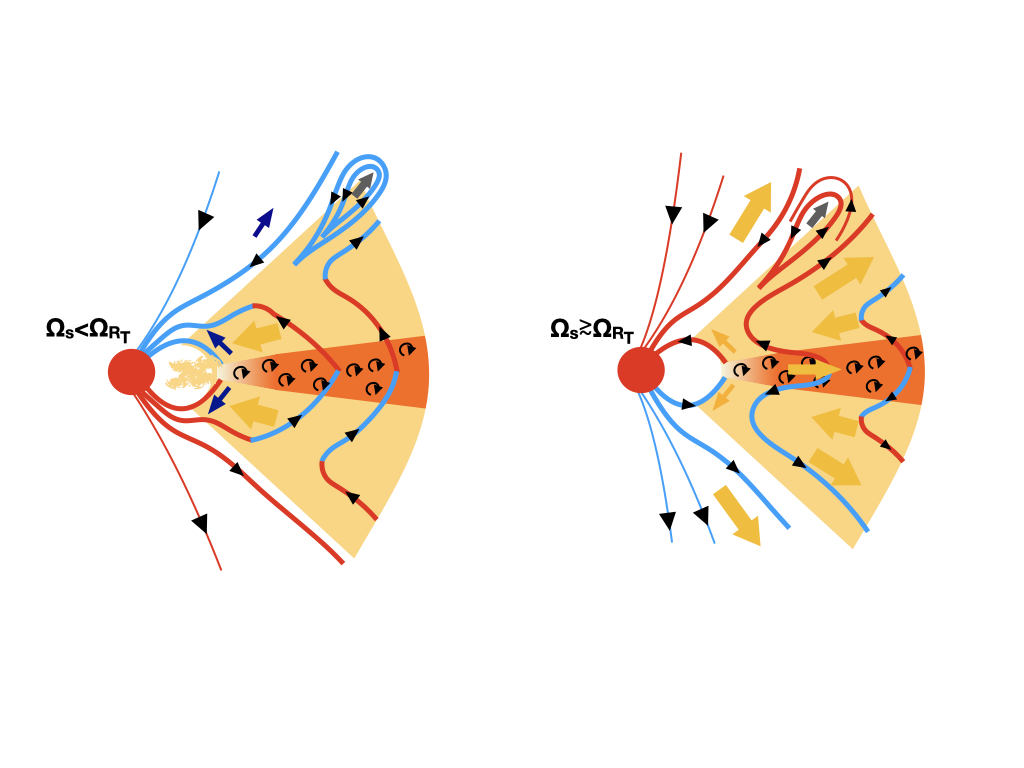}
\figcaption{Velocity and magnetic structure at different times (top to bottom panels, the fourth row: the end of the simulations) for simulations with $R_c=2 R_0$ (left two columns) and $R_c=0.5 R_0$ (right two columns). White contours represent regions where $\overline\beta\equiv2\langle P\rangle/(\langle B_{r}\rangle^2+\langle B_{\theta}\rangle^2+\langle B_{\phi}\rangle^2)=1$, and green contours indicate where $(\langle v_{r}\rangle^2+\langle v_{\theta}\rangle^2+\langle v_{\phi}\rangle^2)\cdot\langle\rho\rangle/(\langle B_{r}\rangle^2+\langle B_{\theta}\rangle^2+\langle B_{\phi}\rangle^2)=1$. Streamlines in the velocity panels represent the poloidal velocity,
while those  in the magnetic field panels represent the poloidal magnetic field. The bottom panels show schematic diagrams of magnetic fields and flow structures for slow and fast rotators. Red and blue curves are magnetic field lines with positive and negative $B_{\phi}$ components, respectively. Arrows indicate the
 poloidal flow direction, with colors (yellow or purple) representing positive and negative $v_{\phi}$ in the region. 
 \label{fig:evolution}}
\end{figure*}

\begin{figure*}[t!]
\centering
\includegraphics[trim=0mm 0mm 0mm 10mm, clip, width=6.in]{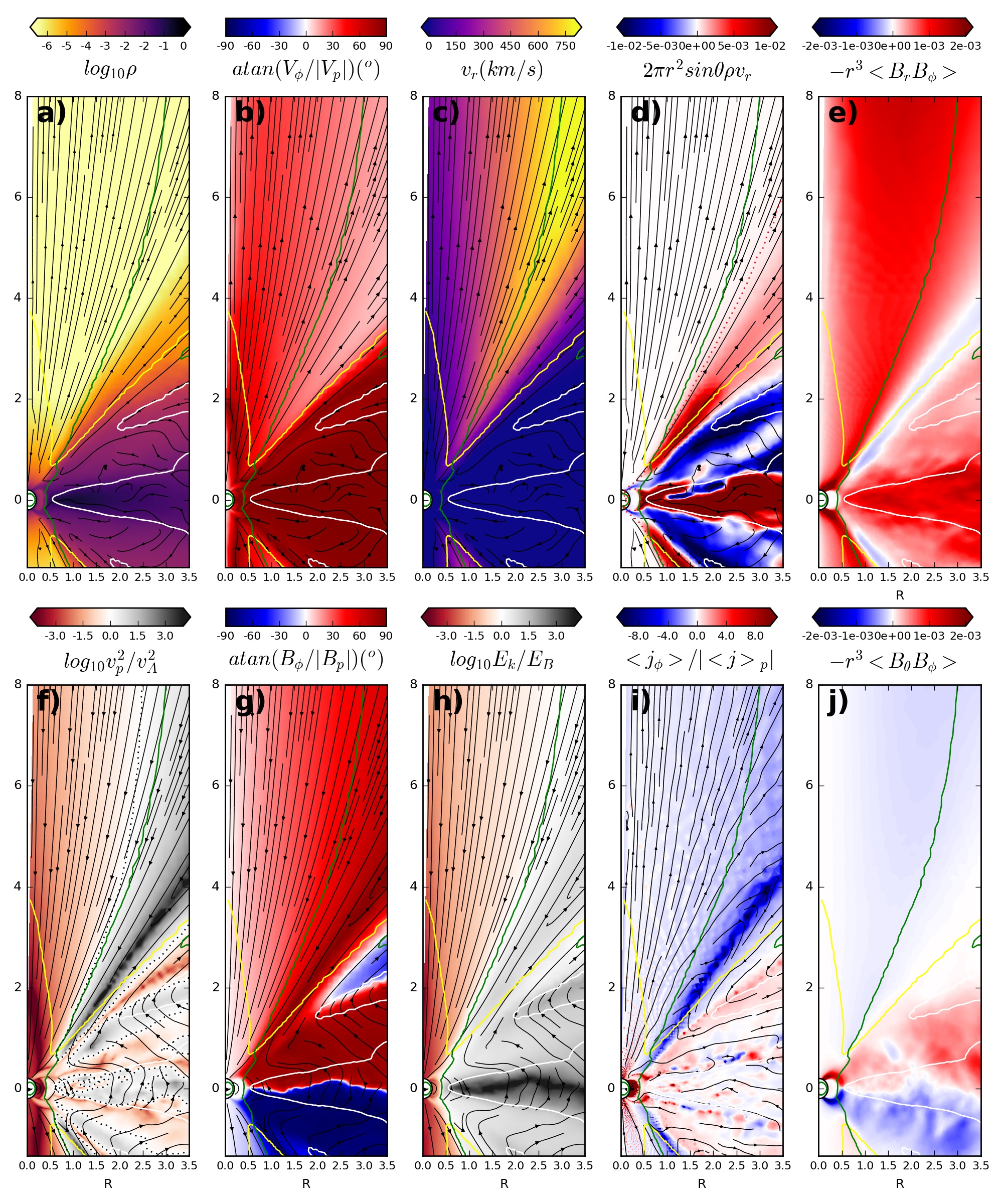}
\figcaption{ Time and azimuthally averaged density, velocity angle, radial velocity, mass flux, $r-\phi$ Maxwell stress (upper panels), and poloidal velocity, magnetic field angle, energy ratio, electric current, and $\theta-\phi$ Maxwell stress (lower panels) for the $R_c=0.4 R_0$ case. All primitive quantities (velocity and magnetic fields), $\rho v_{r}$, and stresses are averaged over 50 snapshots for the last 5 orbits. The white, green, and yellow curves in all panels indicate where $\beta$=1, $E_k=E_B$, and $v_r=\sqrt{2GM_*/r}$, respectively.   Dotted curves in the lower left panel mark where $v_p=v_{A,p}$, while the red dotted curves in the mass flux panel indicate where the passive scalar concentration ($r_{disk}$) is 0.5. 
Streamlines in the upper panels represent the poloidal velocity, while those in the lower panels present the poloidal magnetic fields, except in the electric current panel, where poloidal electric currents are shown. 
\label{fig:outflowavg}}
\end{figure*}

\begin{figure*}[t!]
\centering
\includegraphics[trim=0mm 0mm 0mm 10mm, clip, width=5.in]{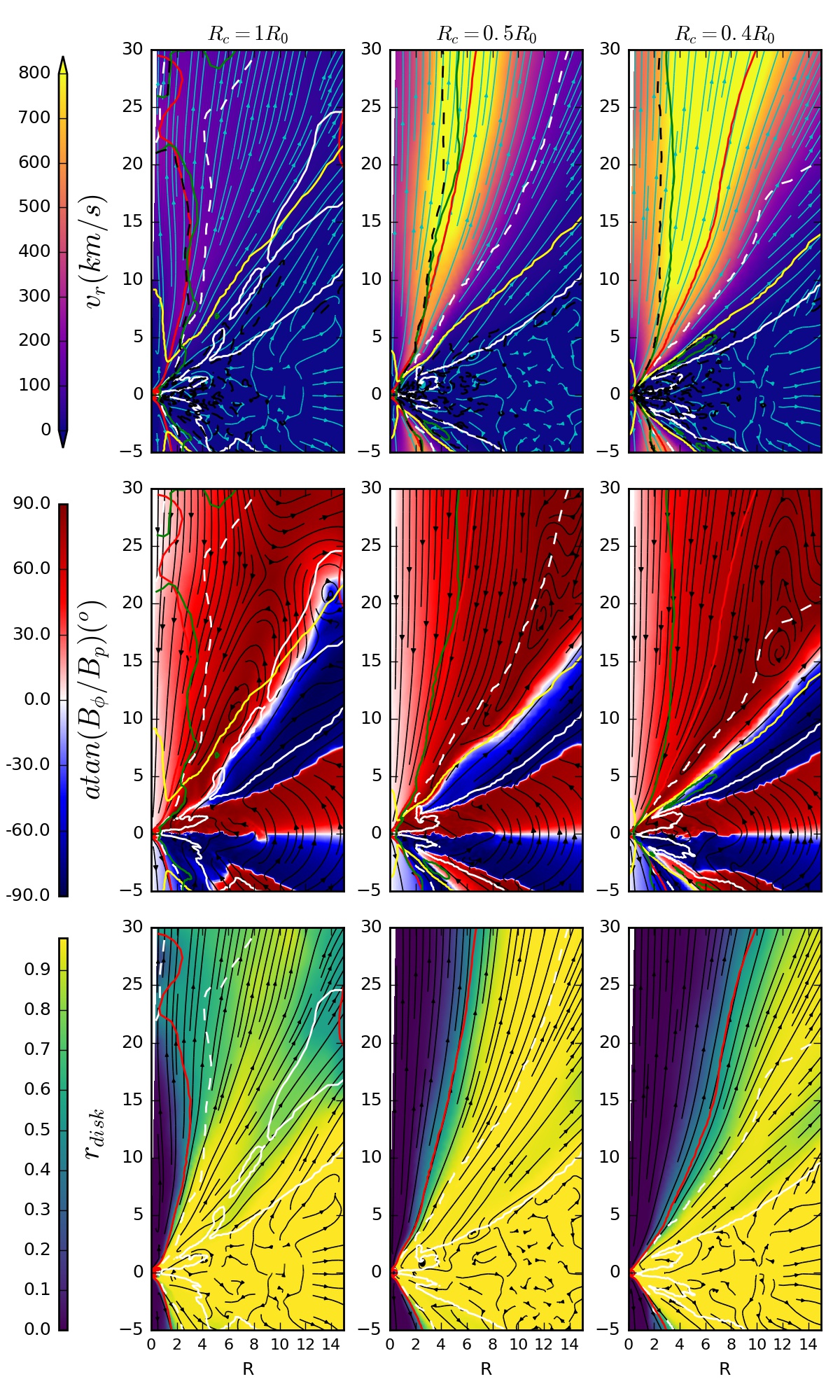}
\figcaption{ Azimuthally averaged radial velocity, magnetic field pitch angle, and disk passive-scalar fraction ($r_{disk}$) for the three cases at the end of the simulation. The white streamlines in the first row represent poloidal velocity, while the black streamlines in the second row represent poloidal magnetic fields. The white, red, green, and yellow solid curves in all panels indicate where $\beta$=1, $r_{disk}=0.5$, $E_k=E_B$, and $v_r=\sqrt{2GM_*/r}$, respectively.  The white dashed curves indicate where $\beta=0.01$.} 
\label{fig:outflow}
\end{figure*}

\begin{figure*}[t!]
\centering
\includegraphics[trim=0mm 5mm 0mm 0mm, clip, width=7.in]{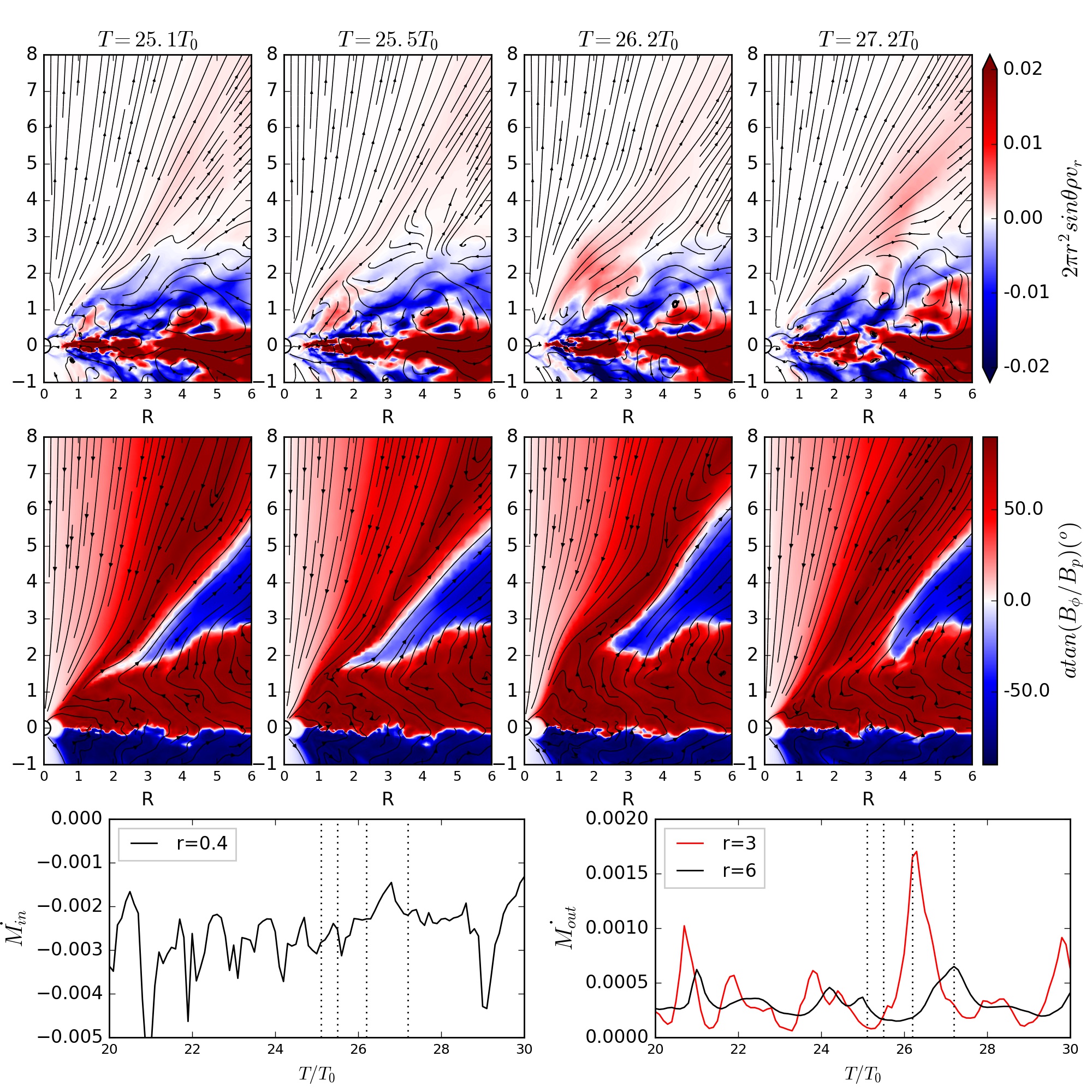}
\figcaption{  Upper two rows: the azimuthally averaged radial mass flux (the first row) and magnetic field pitch angle (the second row) at four different times (from left to right panels). The bottom row: the mass accretion rate (the left panel) and the outflow rate (the right panel) versus time. The four snapshots shown at the top two rows correspond to the times labeled as vertical dotted lines in the bottom panels.
\label{fig:spacetimeoutflow}}
\end{figure*}


While the slow rotators produce weak winds \citep{Zhu2024}, fast rotators could produce much stronger winds. To understand the launching mechanism of the wind, it is insightful to study how the magnetic fields develop over time and why the field configurations differ between the slow and fast rotators, as shown in Figure \ref{fig:evolution}. At the very early stage (0.3 $T_0$ panels), the disk's radial differential rotation stretches the dipole fields within the disk, producing significant $B_{\phi}$ within the disk. However, in the magnetically dominated atmosphere region, the different rotation rates between the star and the disk stretch the fields differently for rotators at different spin rates. For the slow rotator, the faster disk rotation produces negative $B_{\phi}$ in the atmosphere at $z>0$, while for the fast rotator, the faster stellar rotation produces positive  $B_{\phi}$ there. In both cases, the amplified $B_{\phi}$ expands due to increasing magnetic pressure and pushes the atmosphere outwards. Eventually, these magnetic loops open up and leave two sharp $B_r$ transitions: one at the disk surface where $\beta\sim$1 and one at $z\sim R$ (as in the 0.5 $T_0$ panel). Reconnection from the opposite $B_r$ at $z\sim R$ keeps generating magnetic loops that propagate outward (as in the 2 $T_0$ panel). These reconnection events lead to episodic wind.  Meanwhile, MRI starts to develop within the disk and generate turbulence, allowing magnetic fields to diffuse in the disk. Such turbulent diffusion is crucial for maintaining a quasi-steady state for accretion \citep{Ghosh1979II, Zhu2024}. For fast rotators, there is a negative $B_{\phi}$ region sandwiched between the disk and surface $B_r$ transition region. This region results from the disk surface accretion and radial differential rotation \citep{ZhuStone2018}. Surface accretion pinches the fields inwards, and differential rotation stretches the fields azimuthally, leading to $B_{\phi}\cdot B_r<0$. This sandwiched region could also lead to disk wind. However, the outflow's radial speed in this small region hardly reaches the local Keplerian speed. 

To study the wind in more detail, we have averaged quantities for the last 5 orbits in the Rc0p4 case (which has the strongest wind), as shown in Figure \ref{fig:outflowavg}. The disk is clearly separated into three regions: the outflow midplane, the inflow surface, and the outflow wind (Panel d). To ensure that the wind is not produced by the density floor close to the star, we have restarted the simulation and added passive scalar only in the disk region ($|\theta-\pi/2|<40^o$). The passive scalar is then advected into the wind region. The passive scalar concentration ($r_{disk}$, shown in the bottom panels of Figure \ref{fig:outflow}), defined as the ratio between the passive scalar density and the gas density, represents the fraction of the mass in each cell that is originally in the disk. Thus, $(1-r_{disk})$ is the upper limit of the density contamination fraction from the density floor.  The red curves ($\sim 20^o$ from the pole) in all the panels  label where $r_{disk}=0.5$,  and thus can also be considered as the separation between the disk and stellar winds. Most outflowing material (including part of the very fast outflowing region) originates outside this region and thus is from the disk.

Overall, we define the region above the yellow curve in Figure \ref{fig:outflowavg} ($\sim$50$^o$ from the pole) as the wind region, which is unbound from the central star. This wind region is separated into the lower part that originates from the disk (disk wind) and the part within $\lesssim 20^o$ from the pole that originates from the star (stellar wind). We can also see that the region slightly below the yellow curve is also outwards moving although at a speed that will not escape the system. These are ``failed wind'' as in \cite{Romanova2018,Takasao2022}.


The wind is driven by field inflation and magnetic reconnection. Most wind occurs at $z\sim R$ where $B_{r}$ changes sign (Panel d and g in Figure \ref{fig:outflowavg}). The magnetic reconnection between the reversed field lines drives the highly variable wind. A strong azimuthal electric current is produced from the reversal of $B_r$ fields (Panel i).  This current generates a strong radial Lorentz force ($\mathbf{J}\times \mathbf{B}$) that drives the wind. The flow gains high speed, reaching more than 500 km/s (Panel c). The $z\sim R$ region outside $R_T$ has strong azimuthal fields (Panel g) but its magnetic energy is smaller than the kinetic energy (Panel h), indicating that most acceleration occurs close to the star. 

This wind launching by reconnection is quite different from the magnetocentrifugal wind in disks with net vertical fields where $B_{r}$ and $B_{\phi}$ change signs simultaneously. Here, reconnection occurs where $B_r$ changes sign but $B_{\phi}$ keeps the same sign, or vice versa. This wind driven mechanism is more similar to the ``field inflation'' proposed by \cite{Aly1990, Lovelace1995, Uzdensky2002}. 
To illustrate wind launching by magnetic field reconnection near the magnetospheric truncation radius, we select four time snapshots following a single wind launching event, as shown in   Figure \ref{fig:spacetimeoutflow}.  Depicted in the leftmost panel in the middle row, reconnection occurs around $r\sim R_c$ where $B_r$ changes sign, generating an outflow in that region (light red region in the leftmost panel in the upper row). Over time, this outflow propagates outward. In the third column, the outward-moving material is located at $r\sim 3 R_0$, while in the fourth column, it moves to  $r\sim 6 R_0$. 
For slow rotators, magnetic reconnections also occur. However, the radius where $B_r$ changes sign is larger and the wind is much weaker (Figure \ref{fig:evolution} and the movie in Figure \ref{fig:outflow}). It is possible that, due to slow rotators' weaker twists and stronger turbulence from the interchange instability, turbulent diffusion in the inner region reconnects the fields before large-scale magnetic fields can twist, inflate, and reconnect.  Note that the wind has opposite $B_{\phi}$ fields for slow and fast rotators.  

\begin{figure*}[t!]
\includegraphics[trim=8mm 5mm 0mm 0mm, clip, width=7.7in]{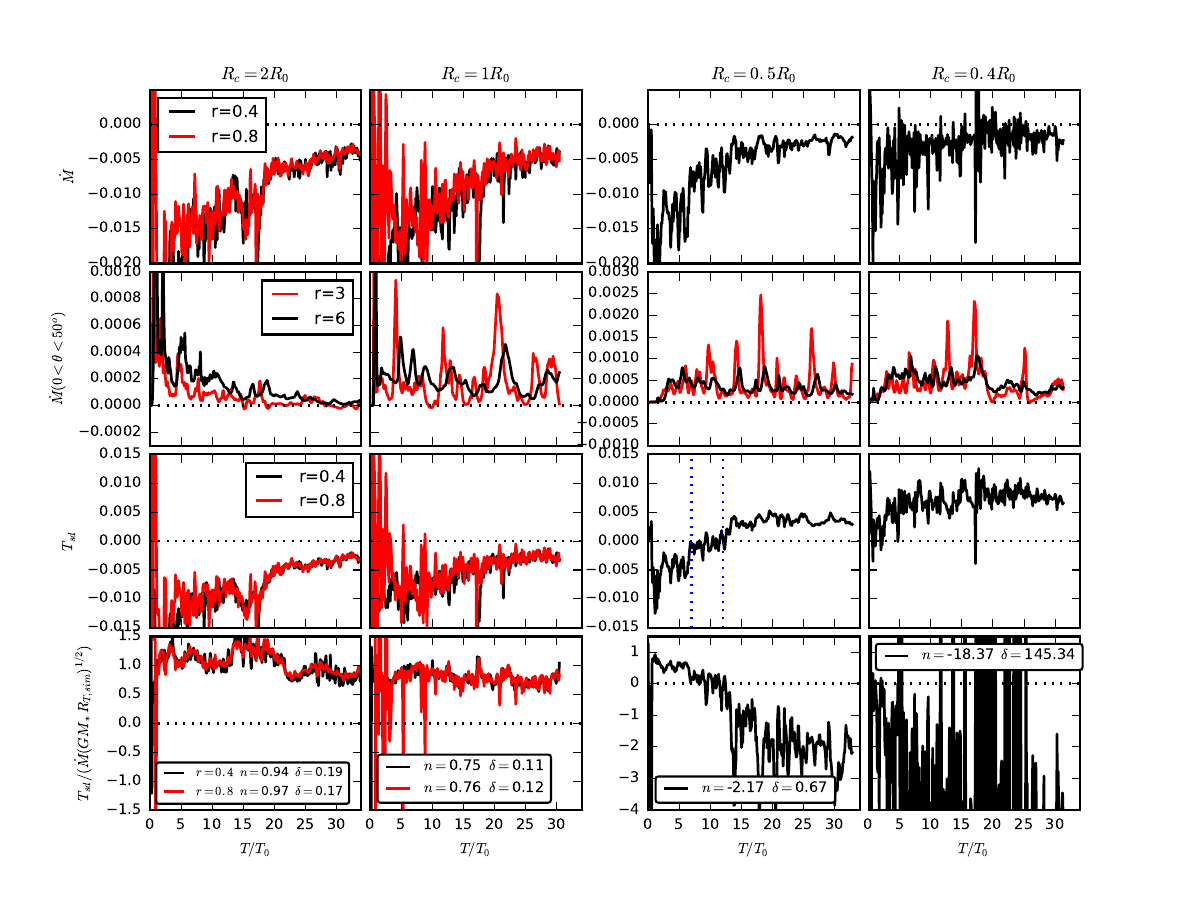}
\figcaption{ Time evolution of the mass accretion rate, mass outflow rate, star-disk torque, and $n$ defined in Equation \ref{eq:nvalue} for all four cases. The red and black curves represent measurements at different  radii. The red curves, corresponding to $r=0.8$, are not shown in the right two panels due to large fluctuations caused by highly variable winds. The mean and standard deviation of $n$ for  $t>$17 $T_0$ are listed in the bottom panels. In the Rc0p5 case, we consider that the disk is in the equilibrium spin state during the timespan between the two vertical blue lines (7 and 12 $T_0$).
\label{fig:mdottime}}
\end{figure*}

\begin{table*}[t!]
\caption{Corotation radius,  accretion rate at $r=0.4 R_0$, wind mass loss rate at $r=6 R_0$ (half disk), wind-to-accretion ratio, truncation radius calculated using Equation \ref{eq:RTcode}, truncation radius measured in simulations, and star-disk torque.  The mean and standard derivation values are taken from the last 5 $T_0$ in Figure \ref{fig:mdottime} and Figure \ref{fig:truncation}. The third row, showing the star-disk in the equilibrium spin state, uses quantities from time 7 to 12 $T_0$ in Rc0p5. Similarly, the first row for the thin disk uses quantities from time 8.5 to 15 $T_0$. The $\dot{M}_{wind}$ is defined as twice the value of $\dot{M}_{out}(0-50^o)$. Note that all the ratios in the table are first calculated at each time snapshot with an interval of 0.1 $T_0$ and then averaged. Thus, the values in the fourth column is not simply twice the ratio between the third and second columns.  }
\small\centering
\begin{tabular}{ c | c | c | c | c  | c | c | c}
 \hline
 $R_c/R_0$ & $\dot{M}_{acc}\times 10^3$ & $\dot{M}_{out}(0-50^o)\times 10^3$ & $\dot{M}_{wind}/\dot{M}_{acc}$ & $R_{T,ana}/R_0$ & $R_{T}/R_0$ & $\frac{T_{sd}}{\dot{M}\sqrt{GM_*R_{T,ana}}}$ & $\frac{T_{sd}}{\dot{M}\sqrt{GM_*R_{T}}}$ \\ 
 \hline
  \hline
2 & -4.37$\pm$1.02 & 0.026$\pm$0.010 & -0.0121$\pm$0.0034  & 0.6004$\pm$0.0376 & 0.7625$\pm$0.0355 & 0.9011$\pm$0.0837 & 0.8004$\pm$0.0856\\  
1 & -4.74$\pm$0.82 & 0.165$\pm$0.069 & -0.0736$\pm$0.0263  & 0.5846$\pm$0.0296 & 0.6119$\pm$0.0301 & 0.7390$\pm$0.0865 & 0.7228$\pm$0.0902\\
0.5 & -7.06$\pm$1.97 & 0.422$\pm$0.084 & -0.1289$\pm$0.0300  & 0.5270$\pm$0.0456 & 0.3926$\pm$0.0186 & 0.0679$\pm$0.2279 & 0.0723$\pm$0.2708\\
0.5 & -2.25$\pm$0.63 & 0.253$\pm$0.056 &-0.2482$\pm$0.0817  &0.7284$\pm$0.0530 & 0.5168$\pm$0.0208 & -1.9604$\pm$0.4518 &-2.3421$\pm$0.6063\\
0.4 & -1.96$\pm$1.01 & 0.269$\pm$0.065 & -0.4016$\pm$0.3086  & 0.7948$\pm$0.1580 & 0.4218$\pm$0.0153 & -6.1048$\pm$5.6290 & -9.0377$\pm$10.598\\
\hline
\multicolumn{2}{l|}{Thin disk} & $\dot{M}_{out}(0-60^o)\times 10^3$ \\
\hline
0.5 & -7.18$\pm$2.72 & 0.38$\pm$0.114 & -0.1235$\pm$0.0481  & 0.5295$\pm$0.0568 & 0.3729$\pm$0.0246 & 0.017$\pm$0.337 & 0.0079$\pm$0.4096\\
0.5 & -4.07$\pm$1.37 & 0.292$\pm$0.071 &-0.162$\pm$0.0546  &0.6204$\pm$0.0621 & 0.3987$\pm$0.0209 & -0.8103$\pm$0.489 &-1.0334$\pm$0.6649\\
 \hline
\end{tabular}
\label{tab:summary}
\end{table*}

\begin{figure*}[t!]
\centering
\includegraphics[trim=0mm 5mm 0mm 3mm, clip, width=6.5in]{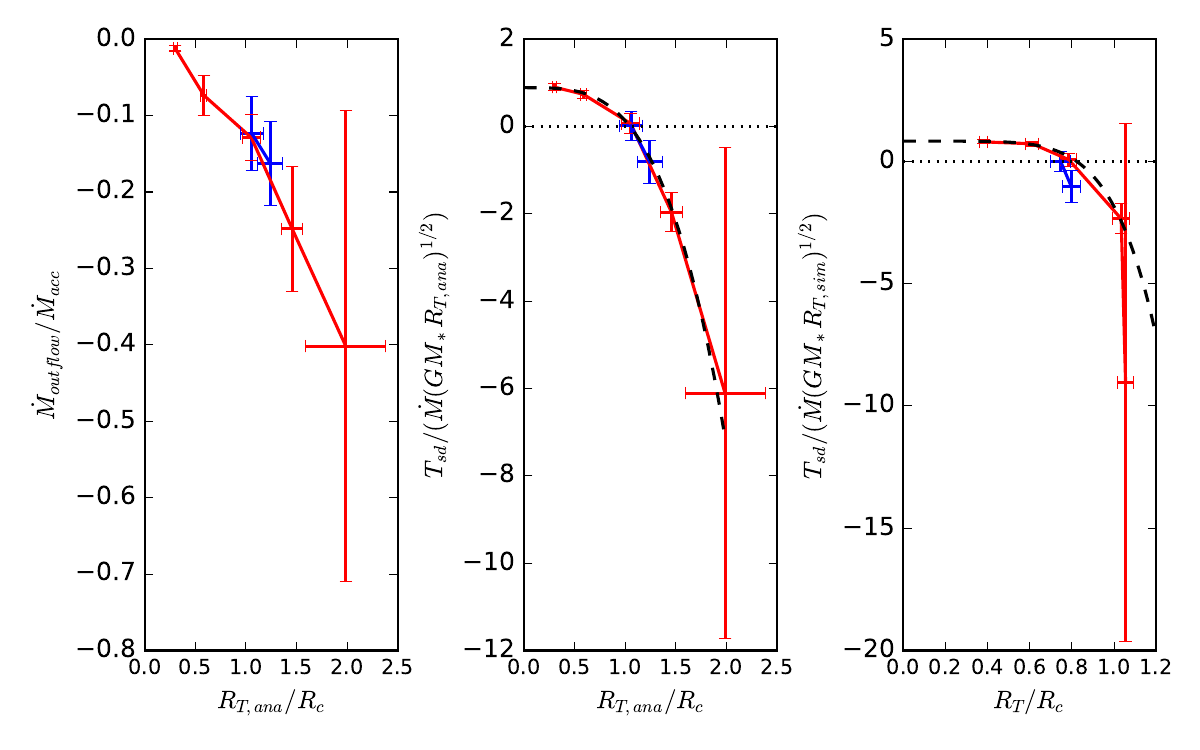}
\figcaption{ Left panel: The ratio between the wind mass loss rate at $r=6 R_0$ to the inward accretion rate at $r=0.4 R_0$ for different cases. Middle and right panels: The stellar spin torque ($T_{sd}$) for different cases, normalized by $R_{T,ana}$ or $R_{T}$. The dashed curves represent the fits from Equations \ref{eq:fitana} and \ref{eq:fitsim}. All data are from Table \ref{tab:summary}.  
\label{fig:outtorque}}
\end{figure*}

To study how the accretion and wind vary with time, we integrate the mass flux at $r=0.4 R_0$ and 0.8$R_0$ over the 4$\pi$ sphere to derive the accretion rate, and the mass flux at $r=3 R_0$ and 6$R_0$ above the disk surface ($\theta<50^o$) to derive the wind outflowing rate, as shown in the top two panels of Figure \ref{fig:mdottime}.  The accretion rates decrease over time as the inner disks are depleted due to accretion. Towards the end of the simulations, the disks reach quasi-steady accretion over a moderately wide radial range, as also shown in Figure \ref{fig:oned}. However, the wind is highly variable, similar to the non-rotator case in \cite{Zhu2024}. The variability is likely due to the field inflation and reconnection. The wind is less variable at $r=6$ compared with $r=3$, indicating that the variability is produced close to the star and damps while propagating outwards.

The average mass accretion rates, the wind outflowing rates, and their standard derivations from the last 5 $T_0$ are given in Table \ref{tab:summary}. The ratios between the wind outflowing rates and the accretion rates are also presented in Table \ref{tab:summary} and Figure \ref{fig:outtorque}. This ratio increases dramatically with the stellar spin, from 1.2\% for slow rotators to 40\% for our fastest rotator.  We note that, although our fastest rotator's $R_{T,ana}$ is twice of $R_{c}$, its $R_{T}$ is merely larger than $R_c$. Thus, our fastest rotator case is not in the strong ``propeller'' regime where the outflow rate can be much higher than the accretion rate \citep{Romanova2018}.  Overall, the wind's sensitive dependence on the stellar spin may have observational implications that will be discussed in \S 5.4.

\subsection{Magnetospheric Truncation Radius}
Magnetospheric truncation radius  can be defined in various ways (see \citealt{Takasao2022}). Following \cite{Zhu2024}, we define it as the radius at the disk midplane where $E_k=E_B$, denoted as $R_T$. We find that, for both slow and fast rotators, this radius nicely corresponds to where the velocity and density change sharply, as shown in Figure \ref{fig:mid1d}. Thus, our $R_{T}$ reasonably represents the inner edge of the disk. 

\begin{figure*}[t!]
\includegraphics[trim=0mm 70mm 0mm 3mm, clip, width=7.in]{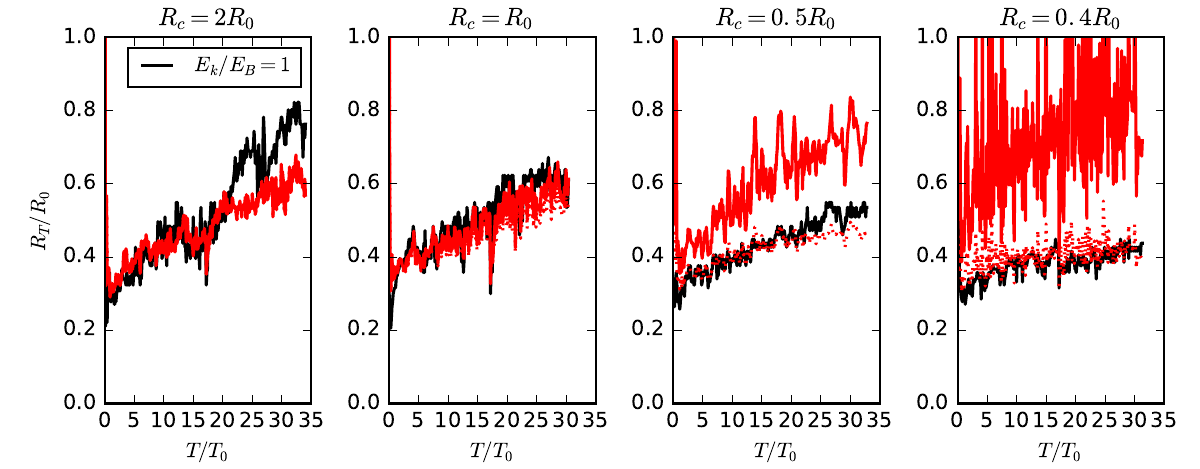}
\figcaption{ Time evolution of the magnetospheric truncation radius for different cases. The red curves are derived using the disk's accretion rate at $R=0.4 R_0$ and Equation \ref{eq:RTcode}. The solid black curves represent direct measurements from the simulations at the midplane where $E_k=E_B$. The dotted curves are calculated using the empirical Equation \ref{eq:emp}. 
\label{fig:truncation}}
\end{figure*}

The truncation radius can also be estimated analytically using Equation \ref{eq:RTcode} as $R_{T,ana}$. Figure \ref{fig:truncation} shows that  $R_{T,ana}$ and $R_{T}$ agree very well for slow rotators (as in \citealt{Zhu2024}), indicating that the ram pressure of accretion balances the magnetic pressure.  However, for fast rotators, $R_{T}$ becomes significantly smaller than $R_{T,ana}$ and it is closer to $R_{c}$. It seems that the rotating stellar dipole allows disk material moving inside of $R_{T,ana}$, but stopping around $R_c$ where centrifugal force starts to dominate. The centrifugal force leads to  interesting 2-D flow structure in the region around $R_T$ (Figure \ref{fig:outflowavg}). The surface accretion brings sub-Keplerian disk material to the magnetosphere. While some material is directly channeled to the central star, some moves to the disk midplane and is centrifugally expelled out by the rotating stellar dipole, leading to midplane outflow. 

Finally, we fit  $R_{T}$ empirically with
\begin{equation}
R_{T,emp}=R_{T,ana}\left(1-e^{-1.5 R_c/R_{T,ana}}\right)\,,\label{eq:emp}
\end{equation}
which has the limit of that, for slow rotators, $R_{T,emp}\sim R_{T,ana}$, while, for extremely fast rotators, $R_{T,emp}\sim 1.5 R_c$. Equation \ref{eq:emp} is shown in Figure \ref{fig:truncation}, fitting $R_{T}$ reasonably well for both slow and fast rotators. 
 Our results are different from some previous 2.5-D viscous/resistive simulations in the propeller regime \citep{Romanova2004b} where the truncation radius can be one order of magnitude larger than  the corotation radius. Our truncation radius is quite close to the corotation radius (more similar to the 3-D MHD simulations by \citealt{Takasao2022}). 
This could be due to efficient field diffusion in MRI turbulent disks. This explanation is supported by
previous 2.5-D simulations having MRI and large diffusivity coefficient at $R_T$ \citep{Romanova2018}, whose truncation radius is closer to the corotation radius. 

Traditionally, the $R_T/R_c>1$ is the criterion for the propeller regime \citep{Illarionov1975}. The propeller regime is associated with several characteristics, including inhibited accretion by the centrifugal barrier, strong outflow due to the centrifugal acceleration, and the spin down of the central star. In this article, we limit the use of the term propeller regime to refer our fast rotators due to several reasons. First, the shown difference between $R_{T,ana}$ and $R_{T}$ makes the definition of the magnetospheric truncation radius ambiguous. Second, the transition of some disk features, such as an increasing outflow rate with a faster stellar spin, is gradual (Figure \ref{fig:outtorque}). Third, the predicted feature of the inhibiting accretion is not observed in our simulation, as already pointed out by \cite{Romanova2004b}. Overall, considering these ambiguities, we discuss our simulations case by case, and distinguish them by using general terms such as slow and fast rotators.

\subsection{Torque on the Star and Disk Accretion Structure}

\begin{figure*}[t!]
\includegraphics[trim=0mm 0mm 0mm 0mm, clip, width=7.in]{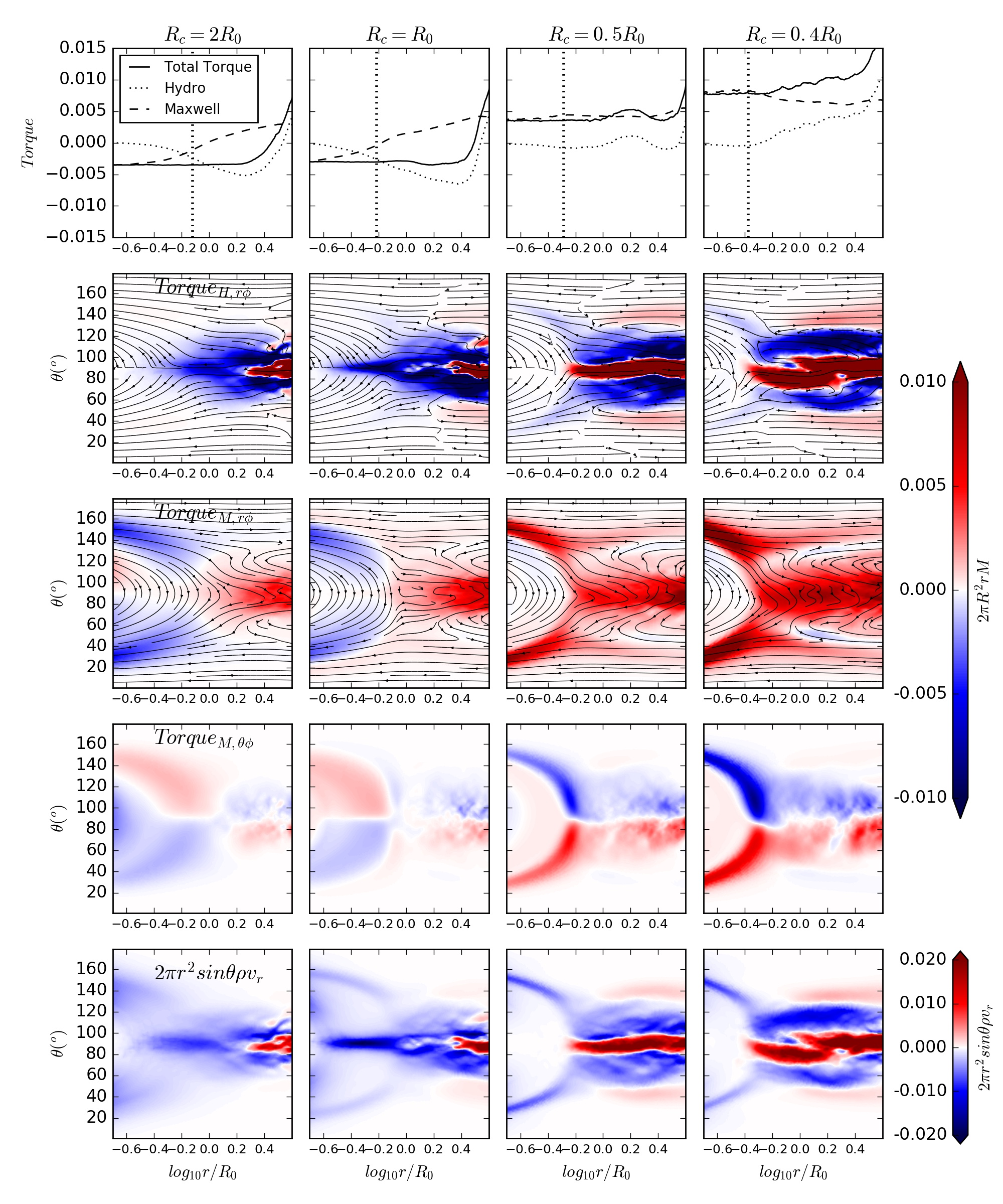}
\figcaption{ Top panels: Integrated torque over spheres at different radii. Second, third, and fourth rows: Hydrodynamical and magnetic
components of the torques along the $\theta$ direction as a function of radius. Bottom panels: Radial mass flux with radius. The streamlines in the second and third rows represent the poloidal velocity and magnetic fields, respectively.
The torques are averaged using 100 snapshots over the last 10 $T_0$.
\label{fig:torque2d}}
\end{figure*}

Magnetospheric accretion couples disk accretion with stellar spin evolution because the total angular momentum of the star and disk is conserved.
The torque between the star and the disk is from two contributions (Equation \ref{eq:amestar}): the magnetic stress -$B_r B_{\phi}$ and the hydrodynamical stress $\rho v_r v_{\phi}$ that includes both the turbulent stress ($\rho v_r \delta v_{\phi}$, or Reynolds stress)  and the angular momentum carried by the accreting material ($\rho v_r \langle v_{\phi}\rangle$). We integrate the total torque over the spheres at $r=0.4 R_0$ and 0.8$R_0$ for slow rotators, and over the sphere at $r=0.4 R_0$ for fast rotators. These values are plotted over time in the bottom two rows of Figure \ref{fig:mdottime}. Note that we define the torque $T_{sd}$ to express the disk's angular momentum change.  Thus, the star's angular momentum change is the negative of this torque, and a positive $T_sd$ means that the star will spin down. 
Dividing the torque by $\dot{M}(GM_* R_{T})^{1/2}$ yields the dimensionless torque parameter ($n$):
\begin{equation}
n\equiv\frac{T_{sd}}{\dot{M}(GM_* R_{T})^{1/2}}\,.\label{eq:nvalue}
\end{equation}
For all our simulations, the measured mean values and standard deviations of $n$ for $t>17 T_0$ are shown in the bottom panels of Figure \ref{fig:mdottime}, while values for the last 5 orbits are listed in Table \ref{tab:summary}. For the slowest rotator, $n\sim$1, similar to the non-rotator case in \cite{Zhu2024}. For our fastest rotator case, $n\sim$-10, indicating the system is well into the so called ``propellor regime''. 

The stress distribution in the disk determines not only the star-disk coupling but also the disk's internal flow structure (e.g., wind and accretion).
When the inner disk reaches a steady state, the spherically integrated torque remains constant with radius (Equation \ref{eq:torqueconst}, top row in Figure \ref{fig:torque2d}). However, hydrodynamical and Maxwell stresses contribute differently at different radii.
Furthermore, the stress is not uniform  in the $\theta$ direction.  Different stress components in the $r-\theta$ plane are shown in Figure \ref{fig:torque2d}. Near the stellar surface (e.g. $r$=0.4$R_0$), the accretion disk twists the stellar magnetic fields, which torque the star while channeling the accretion flow. For slow rotators, the magnetic fields are dragged forward by the faster rotating disk,  exerting a torque that spins up the star and spins down the disk (negative values in the left two Torque$_{M,r\phi}$ panels). Meanwhile, the infalling gas that follows the forward-pitched field lines rotates in the opposite direction from the disk's rotation, exerting a small hydrodynamical stress to spin down the star and spin up the disk (shown as positive values at $\sim$30$^o$ and 140$^o$ in the leftmost Torque$_{H,r\phi}$ panel)\footnote{To plot quantities on the same color scale, the positive hydrodynamical torque is not prominent in the leftmost panel at the second row of Figure \ref{eq:torqueconst}. Clearer color plots are shown in the middle panels of Figures 19 and 20 in \cite{Zhu2024}. }. The negative hydrodynamical stress due to the intruding filaments at the midplane is also evident. For faster rotators, the magnetic fields are dragged backward by the relatively slower rotating disk, exerting a magnetic torque to slow the stellar spin. On the other hand, the gas accretion along the stellar field lines adds positive angular momentum to the star and decreases the angular momentum of the disk (shown as negative values within the magnetosphere in the rightmost Torque$_{H,r\phi}$ panel of Figure \ref{fig:torque2d}).  

Farther from the star, the flow is not in the force-free regime, and hydrodynamical stress becomes increasingly important. 
For slow rotators, hydrodynamical stress surpasses magnetic stress at the magnetospheric truncation radius (top row in Figure \ref{fig:torque2d}).  
In the outer disk, positive Maxwell stress drives inwards accretion which produces negative hydrodynamical stress.  For fast rotators,  strong $T_{\theta\phi}$ stress ($Torque_{M,\theta\phi}$ panels) drives midplane outflow (bottom panels). The hydrodynamical torque from
 midplane outflow offsets the torque from the surface accretion, keeping the hydrodynamical torque small or positive until much larger radii.  Notably for fast rotators, while $\int\rho v_{r} v_{\phi} dS$ is small or positive, $\int \rho v_r dS$ is negative and the disk still accretes inwards over a broad radial range (Figure \ref{fig:oned}). This occurs because $v_{\phi}$ is significantly sub-Keplerian in the surface inflow region. For fast rotators, the wind also carries a moderate amount of angular momentum away.
 
Overall, the entire domain can be divided into three regions: one with closed stellar field lines around the star, one with open field lines anchored on the star, and one with open field lines threading through the disk. With only initial stellar dipole fields, the open field lines are opened by field inflation, separating these two open regions by a current layer at $z\sim R$ high above the disk. Following the open field lines to the disk midplane, the field lines tilt outwards. Turbulent diffusion can transport closed stellar field lines to  $R\sim 2 - 3 R_0$, extending well beyond the truncation radius\footnote{Strictly speaking, some closed field lines are continuously opened and closed by field inflation and reconnection, which act as turbulent diffusion and enable coupling between different disk regions. }. These closed field lines allow stellar rotation to drive midplane outflow. Furthermore, some disk region with open field lines remains influenced by stellar fields through turbulent diffusion and reconnection.  This enables a large disk region to couple with the star, leading to large $n$ values. This differs from previous claims that field inflation limits the star-disk coupling, reducing the spin-up/down efficiency \citep{Matt2005}.
 
\begin{figure*}[t!]
\includegraphics[trim=0mm 0mm 0mm 10mm, clip, width=7.in]{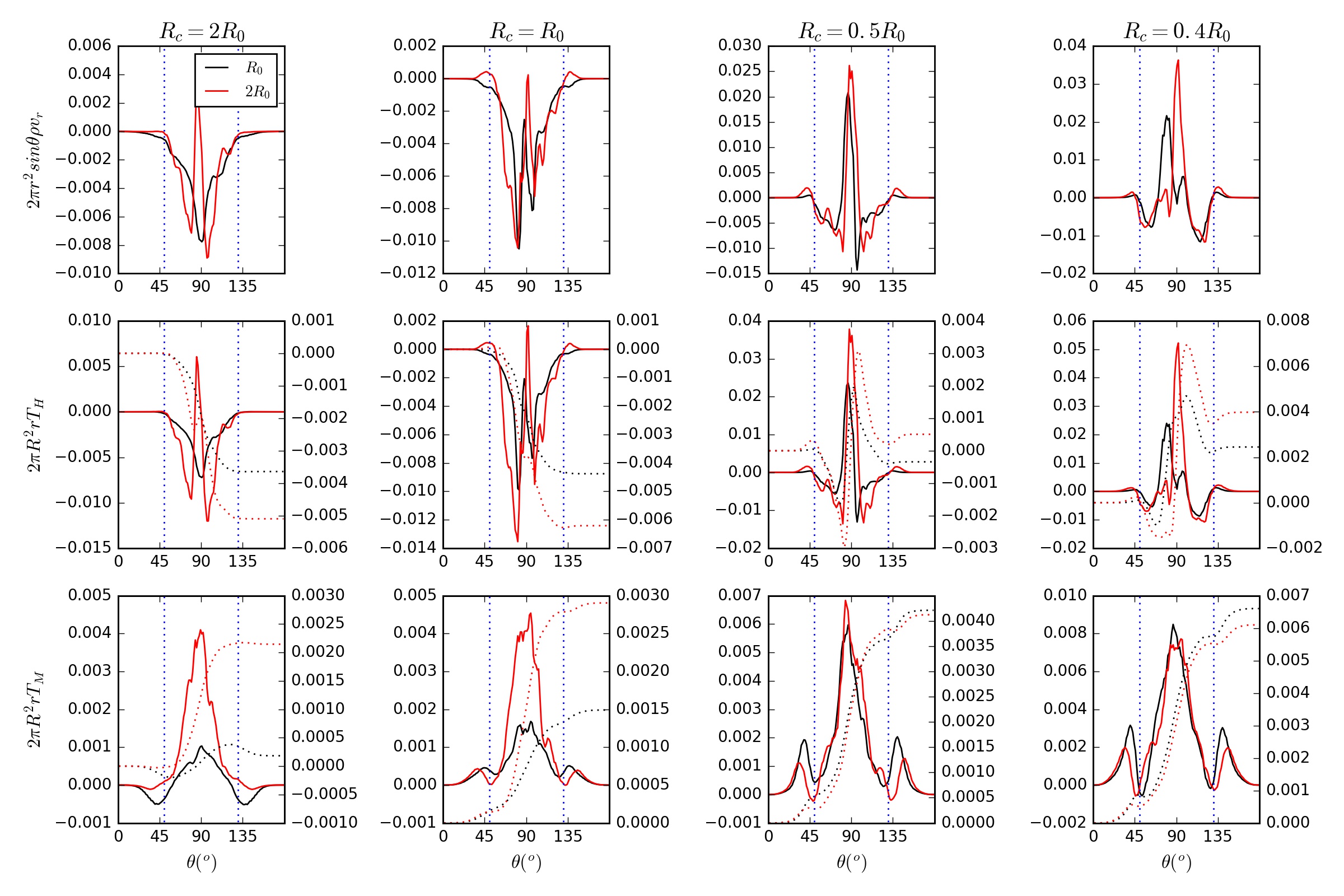}
\figcaption{ Mass flux, and hydrodynamical and magnetic
components of the $r-\phi$ torques (top to bottom rows) along the $\theta$ direction at r=$R_0$ and 2$R_0$ (black and red curves) for different cases (left to right columns). The torques are averaged over 100 snapshots taken during the last 10 $T_0$. These profiles correspond to the vertical cuts shown in Figure \ref{fig:torque2d}. The vertical dotted lines indicate $\theta=50^o$ and $\theta=130^o$, marking the base of the wind. The dotted curves in the middle and bottom rows represent the integrated accumulative torque along the $\theta$ direction with its y-axis shown on the right side of the panels. 
\label{fig:torquetheta}}
\end{figure*} 
 
To quantify the contributions of different regions to the total torque ($T_{sd}$ in Equations \ref{eq:amestar} and \ref{eq:torqueconst}), we plot the mass flux and torques along the $\theta$ direction at $r=R_0$ and $2 R_0$ outside the truncation radius $R_T\sim 0.5 R_0$, as shown in Figure \ref{fig:torquetheta}. For the Rc2 case, the torque in the wind region is negligible, and surface accretion leads to a negative hydrodynamical torque. The disk accretion is driven by positive magnetic torque in the disk. The torque contribution from the wind region (beyond the dotted lines) increases with a faster stellar spin. In the Rc0p5 and Rc0p4 cases, the midplane outflow is  strong enough to offset the torque from surface accretion, making the total hydrodynamical torque nearly zero or even positive. In these cases, the magnetic torque dominates the total torque, with the wind region contributing about 20\% of the total magnetic torque.

We fit the stellar spin-up/down torque with respect to the stellar spin rate, using the measured $n\equiv T_{sd}/\left(\dot{M}\sqrt{GM_* R_{T,ana}}\right)$ and calculated $R_{T,ana}$ listed in Table \ref{tab:summary}:
\begin{equation}
\frac{T_{sd}}{\dot{M}\left(GM_* R_{T,ana}\right)^{1/2}}=0.89-0.76\times \left(\frac{R_{T,ana}}{R_c}\right)^{3.43}\,.\label{eq:fitana}
\end{equation}
If we fit $T_{sd}/\left(\dot{M}\sqrt{GM_* R_{T}}\right)$ against $R_{T}$ instead, we obtain: 
\begin{equation}
\frac{T_{sd}}{\dot{M}\left(GM_* R_{T}\right)^{1/2}}=0.83-2.68\times \left(\frac{R_{T}}{R_c}\right)^{6}\,.\label{eq:fitsim}
\end{equation}
Both fits are shown in Figure \ref{fig:outtorque}, and the fit with $R_{T,ana}$ is better.

The dependence of the stellar spin torque with respect to the stellar spin has been studied both analytically and numerically, with various formulae proposed (e.g. \citealt{Ghosh1979II, Wang1995, Rappaport2004}). While most formulae suggest that $n\sim$1 when $R_c$ is much larger than $R_T$, the $n$ values for disks with $R_T\sim R_c$ differ significantly. Our torque results  also noticeably  differ from recent work by \cite{Takasao2022}, who finds that  $n \sim -0.4$ for their fast rotator ($R_{T}/R_c\sim 1.75$) and $n \lesssim 0.2$ for their slow rotator ($R_{T}/R_c\sim 0.5$). In contrast, our derived $n$ values (Table 1) are much higher: $\sim$ -6 and 0.9 for similarly fast and slow rotators. These differences could be due to the one-sided wind or the stellar wind setup in  \cite{Takasao2022} (more discussion in \S 5.5).

\section{Discussion}

\subsection{The Equilibrium Spin State}
\label{sec:equi}

\begin{figure*}[t!]
\includegraphics[trim=0mm 260mm 0mm 105mm, clip, width=7.in]{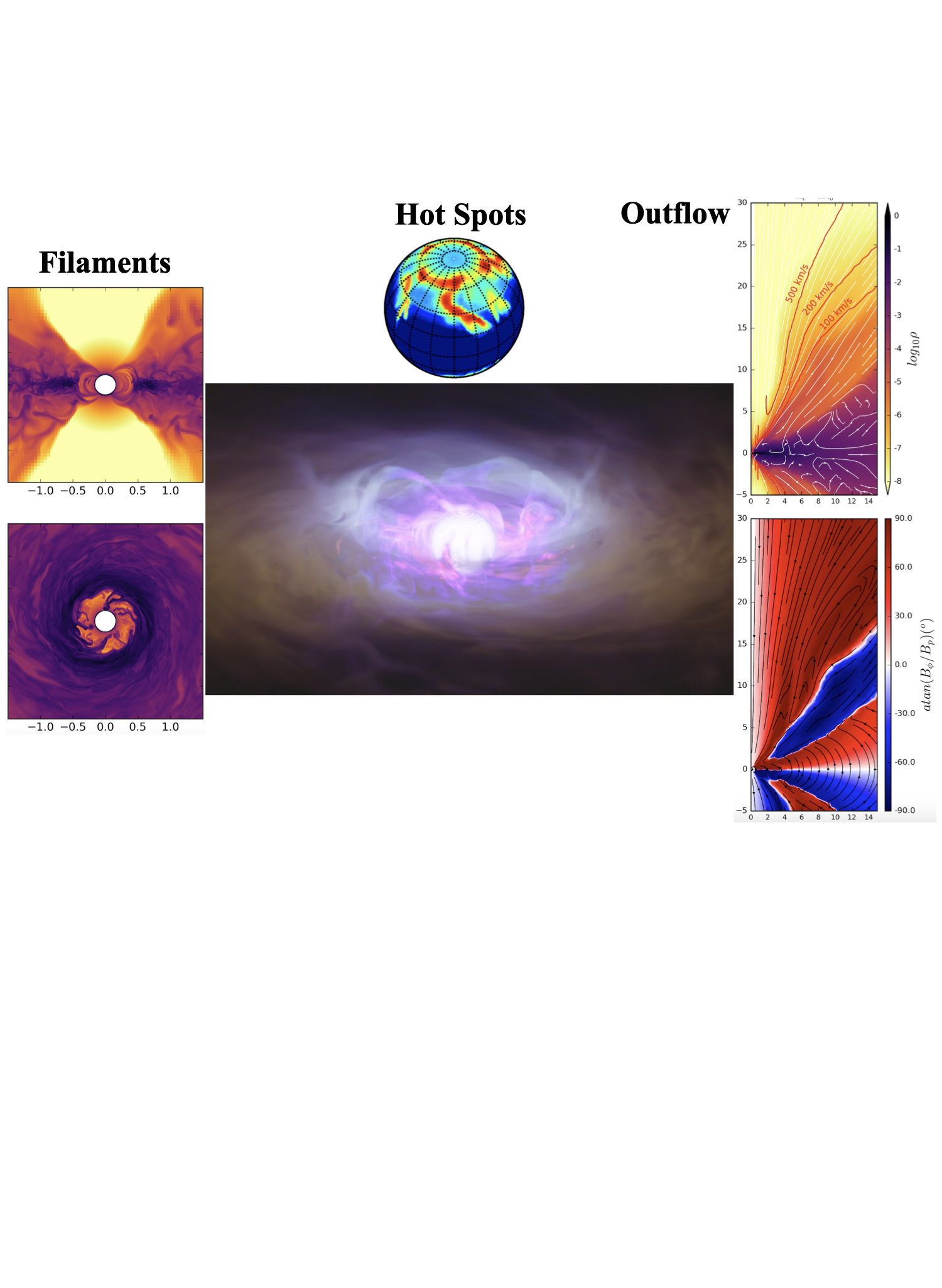}
\figcaption{ Similar to Figures \ref{fig:mid2d}, \ref{fig:sphere} and \ref{fig:evolution}, but for the Rc0p5 run at 10 $T_0$ when the star and disk are in the equilibrium spin state.  The hot spots panel shows the contours of the radial energy flux.
\label{fig:equilibrium}}
\end{figure*}

\begin{figure*}[t!]
\includegraphics[trim=0mm 8mm 0mm 0mm, clip, width=7.in]{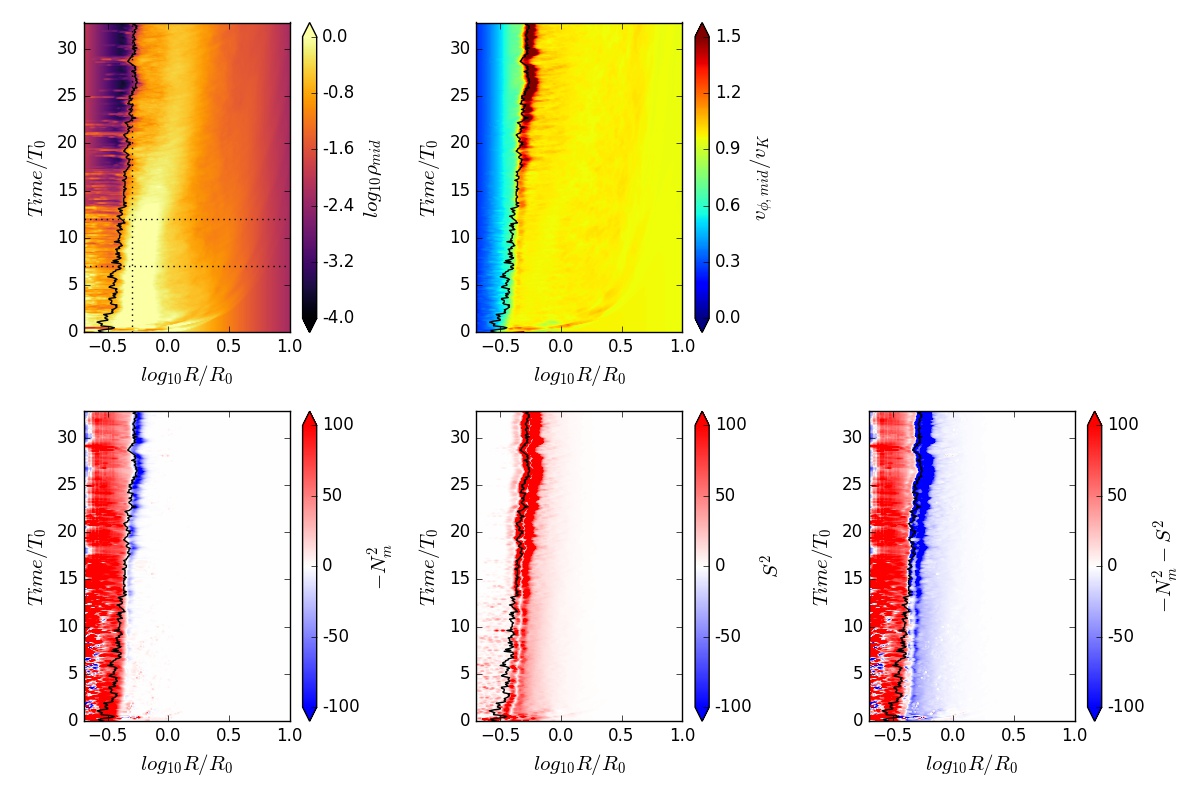}
\figcaption{ Space-time diagrams for the midplane density, $v_{\phi}$, $-N_m^2$, $S^2$, and $-N_m^2-S^2$ for the Rc0p5 case. The black curve in each panel shows $R_{T}$ over time. The vertical dotted line marks $R_c$. The star is in the equilibrium spin state between the two horizontal lines (7-12 $T_0$) in the upper left panel. 
\label{fig:spacetimetrunc}} 
\end{figure*}

\begin{figure*}[t!]
\includegraphics[trim=0mm 250mm 0mm 50mm, clip, width=7.in]{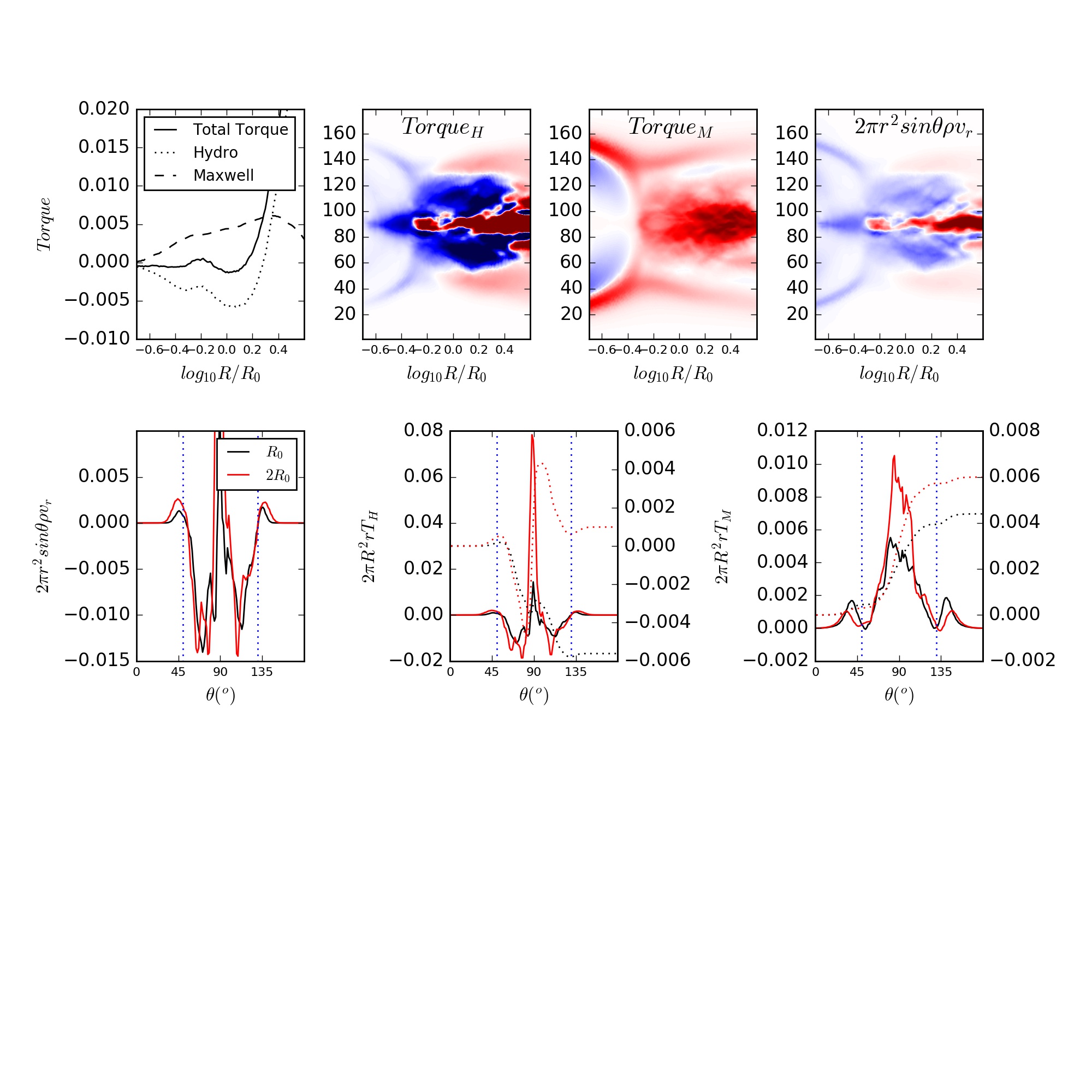}
\figcaption{ Similar to Figures \ref{fig:torque2d}, and \ref{fig:torquetheta}, but for the Rc0p5 run during 7-12 $T_0$, when the star and disk are in the equilibrium spin state. 
\label{fig:equilibrium2}}
\end{figure*}

For stars undergoing magnetospheric accretion, slow rotators ($R_c>>R_T$) will spin up, while fast rotators ($R_c<<R_T$) will spin down. Thus, all accreting stars evolve towards the equilibrium spin state. If the circumstellar disk's lifetime is longer than the spin evolution time (as discussed in the introduction), most stars should reach the equilibrium spin state, making the study of this state highly significant for observational implications.

The equilibrium spin state is achieved in our Rc0p5 simulation during t=7-12$T_0$ (Figure \ref{fig:mdottime}). Since the disk's accretion rate decreases over time in our simulations, $R_T$ moves outward. At earlier times ($t<7 T_0$), 
$R_T$ is too close to the star due to the high $\dot{M}$, causing the star to spin up. At later times ($t>12 T_0$), $R_T$ moves too far away, resulting in the star spinning down. During the equilibrium state, we have
\begin{equation}
R_{T,ana}\sim1.05 R_{c}  \;\;\; {\rm or}  \;\;\;  R_T\sim 0.79 R_c
\end{equation}
(also listed in Table \ref{tab:summary}). This 0.79 ratio in our simulation is slightly above $\sim$0.7  found in \cite{Long2005}, but significantly lower than $\sim$ 1 in \cite{Ireland2021}. However, these comparisons may be affected by differences in simulation setups (e.g., MHD turbulence vs. viscous/resistive simulations, stellar wind treatments,  etc.).

The disk properties during our equilibrium spin state are shown in Figure \ref{fig:equilibrium}. Within the magnetosphere, the presence of filamentary features resemble those of the slow rotator cases more than the fast rotator cases, leading to some hot spots at lower latitudes. Considering that the Rc0p5 case eventually evolves into a fast rotator state without filaments within the magnetosphere, we can study how the ``interchange instability'' is gradually suppressed with increasing $R_T/R_c$ by investigating the disk properties as the disk's accretion rate decreases. The space-time diagrams for various quantities at the midplane are shown in Figure \ref{fig:spacetimetrunc}. The density panel reveals a sharp drop in density within the cavity after $t\sim13 T_0$, indicating suppression of the interchange instability then. As $R_{T}$ retreats and approaches $R_c$, the cavity edge starts to be accelerated by rotating stellar fields and rotate super-Keplerianly. This super-Keplerian rotation results in negative instability growth ($-N_{m}^2<0$) and positive damping from the shear ($S^2>0$), leading to the suppression of the instability. If we consider 13$T_0$ as the critical time when instability is suppressed, $R_{T}\sim 0.85 R_c$ at that time can be considered the critical truncation radius for ``interchange instability''.  Based on Equation \ref{eq:emp}, $R_{T}\sim 0.85 R_c$ corresponds to $R_{T,ana}\sim 1.18 R_C$. Thus, the condition for interchange instability to occur is 
\begin{equation}
R_{T,ana}<1.18 R_c \;\;\; {\rm or} \;\;\; R_{T}<0.85 R_c\,.\label{eq:inter}
\end{equation} Since the equilibrium spin state has $R_{T} \sim 0.79 R_c$ or $R_{T,ana} \sim 1.05 R_c$, the interchange instability should operate during the equilibrium spin state, consistent with our simulation results. 

Outside the magnetosphere, a fast wind reaching up to 500 km/s is launched (Figure \ref{fig:equilibrium}). The disk's accretion rate, wind mass loss rate, and torque during this state are given in Table \ref{tab:summary}. The ratio between the wind  mass loss rate and disk accretion rate is  $\sim$13\%, similar to the observed ratio between jet and disk accretion (e.g., \citealt{Nisini2018}).

During the equilibrium spin state, the torque distribution is more similar to that of the fast rotator case than the slow rotator case  (upper panels in Figure \ref{fig:equilibrium2}).  Within the magnetosphere, the magnetic torque is positive, indicating that the magnetic fields are dragged forward by stellar rotation. The infalling material following the magnetic field lines carries negative angular momentum. The filaments at the midplane also carry significant negative angular momentum. Overall, the negative hydrodynamical torque due to accretion balances with the positive magnetic torque. 
Beyond the truncation radius (e.g. $R_0$ and 2 $R_0$ in bottom panels of Figure \ref{fig:equilibrium2}), the midplane outflow and disk wind carry angular momentum away, producing positive hydrodynamical torque. Meanwhile, the magnetic torque remains universally positive. If we divide the region into the disk region and the wind region, the total torque in the wind region should balances the total torque in the disk region during the steady state, as the sum of these two torques is zero.  If we only consider the magnetic torque, the wind region contributes $\sim$10\% of the total magnetic torque (the fraction of the dotted curve outside the disk region in the lower right panel).

\subsection{Thin Disk}
\label{sec:thindisk}

\begin{figure*}[t!]
\includegraphics[trim=0mm 340mm 10mm 0mm, clip, width=6.5in]{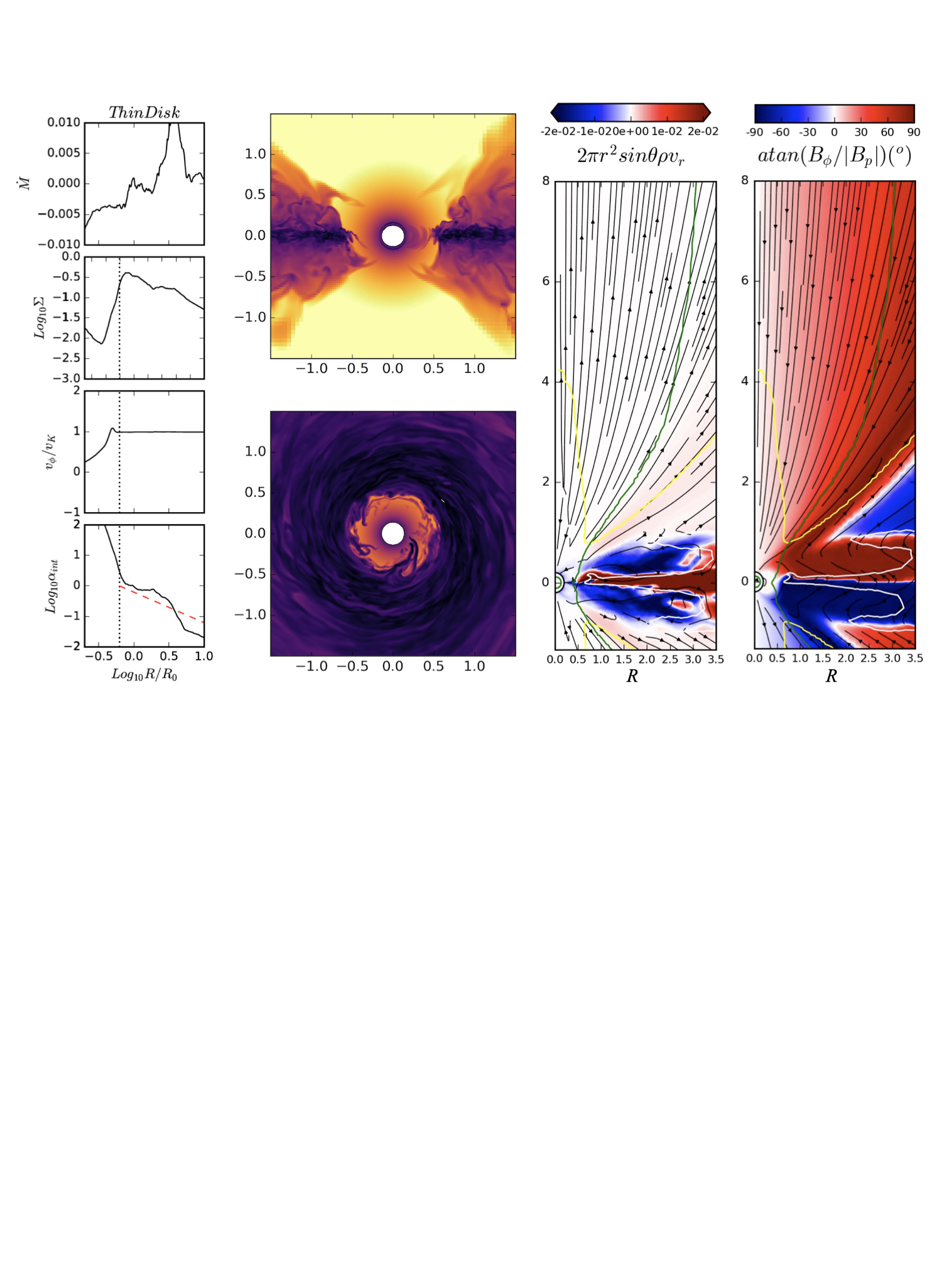}
\figcaption{ Disk structure for the thin disk model ($h$=0.05). The leftmost panels: the mass accretion rate, surface density, midplane azimuthal velocity, and $\alpha_{int}$ at the end of the simulation, similar to Figure \ref{fig:oned}. The vertical dotted line is $R_{T,ana}$. The dashed curve is again $\alpha_{int}=(R/R_{T,ana})^{-1}$.The second column: the poloidal and midplane cuts of the density, similar to Figure \ref{fig:mid2d}. The rightmost two columns: the radial mass accretion rate and field pitch angle that are averaged over the last 5 orbits, similar to Figure \ref{fig:outflowavg}. 
\label{fig:thindisk}}
\end{figure*} 
 
 We choose $R_c=0.5 R_0$ for our thin disk simulation with $h_0=0.05$, which runs until $t=23 T_0$. The star first spins up and then spins down similar to our fiducial Rc0p5 case.  The equilibrium spin state lasts from 8.5 to 15 $T_0$. 

The radial profiles of several quantities are shown in the leftmost column of Figure \ref{fig:thindisk}.  The $\alpha_{int}=(R/R_{T,ana})^{-1}$ profile, which is derived for our fiducial thick disks, fits the thin disk's $\alpha_{int}$ fairly well, indicating that this profile may be insensitive to $h$ and we could use it to estimate the disk's surface density for any given $\dot{M}$. 

The disk's velocity and magnetic structures (the rightmost two columns) are quite similar to our fiducial cases. The surface accretion and midplane outflow are apparent. However, the surface accretion region appears thinner, which extends to $z\sim 0.3 R$ instead of $z\sim R$. Nevertheless, this is still significantly thicker than the disk scale height $\sim 0.05 R$.

To be more quantitative, we have measured the accretion rates, outflow rates, truncation radii, and the stellar torque for both the equilibrium spin state and the final spin down state, given in Table \ref{tab:summary}.  Since the outflow region extends to a lower $z$, we consider the region within $60^o$ from the pole as the outflow region. The outflow efficiency and the stellar torque are plotted in Figure \ref{fig:outtorque}, which agree with the values in thicker disks surprisingly well. This suggests that our outflow-spin relationship and the torque-$R_T$ fit (Equation \ref{eq:fitana}) are universal. Furthermore, the equilibrium state has $R_{T,ana}/R_c\sim 1.06$ and $R_{T}/R_c\sim 0.75$, which are similar to 1.05 and 0.79 in thicker disks (\S \ref{sec:equi}). At the end of the simulations, the disk has $R_{T,ana}/R_c\sim 1.24$ and $R_{T}/R_c\sim 0.80$. Based on our interchange instability condition in Equation \ref{eq:inter}, the disk should have intruding filaments throughout the simulation, which agrees with the simulation.

\subsection{Radial Transport of Magnetic Fields}
\begin{figure}[t!]
\includegraphics[trim=0mm 8mm 0mm 0mm, clip, width=3.5in]{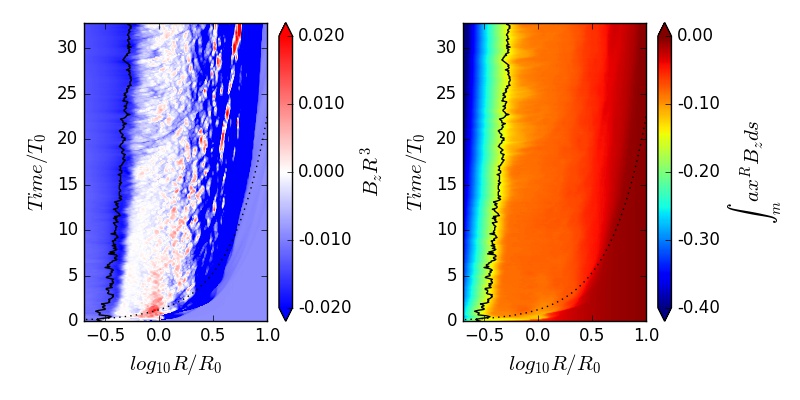}
\figcaption{ Space-time diagrams for the midplane $B_z$ and the integrated midplane $B_z$ flux (integrated from the outer boundary) for the Rc0p5 case.
\label{fig:spacetimefield}}
\end{figure}  

The initial dipole magnetic fields are quickly distorted due to their interaction with the disk. The strong stellar magnetic fields push the disk material outwards. If the influenced region expands at the sound speed:
\begin{equation}
\frac{dR_{c_s}}{dt}=c_s=0.1 v_{\phi_0}\left(\frac{R}{R_0}\right)^{-1/4}\,,
\end{equation}
we can integrate this equation to derive
\begin{equation}
R_{c_s}(t)=\left(\frac{\pi t}{4 T_0}\right)^{4/5}\,.
\end{equation}
This $R_{c_s}(t)$ is plotted in Figure \ref{fig:spacetimefield} as the dotted curves, which agree with  the outer boundary of $B_z$ reasonably well. The small discrepancy at early times is probably due to the neglect of the magnetic pressure in the sound speed calculation.
While $B_z$ is pushed outwards quickly with the disk material, it takes time for MRI turbulence to develop.  Then, the net $B_z$ is reduced in the MRI turbulent region, indicating that magnetic fields are transported outwards by turbulence. This reduction of $B_z$ is similar to \cite{Takasao2022} (Figure 11) and \cite{Zhu2024} (Figure 22). It is unclear if $B_{z}$ will eventually become zero in the disk or the magnetosphere will continue to diffuse $B_z$ out through the disk, which deserves future studies. 

\subsection{Connection with 2-D models}
 \begin{figure*}[t!]
\includegraphics[trim=0mm 8mm 0mm 0mm, clip, width=6in]{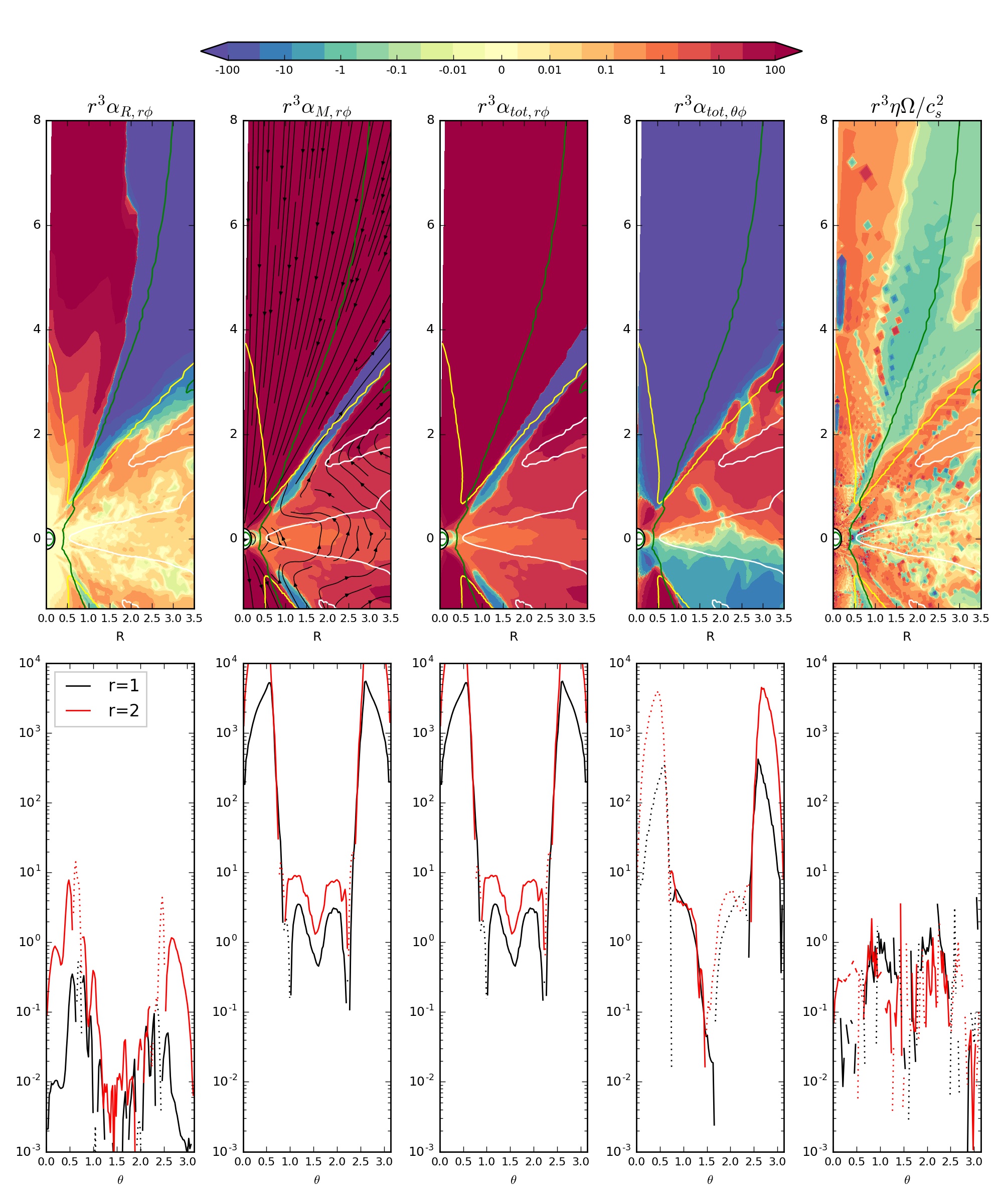}
\figcaption{ Similar to Figure \ref{fig:outflowavg}, but for the $\alpha$ parameters and the resistivity. All quantities have been multiplied by $r^3$ as in Figure \ref{fig:torque2d}. For the Reynolds stress, the azimuthally averaged $v_{\phi}$ has been subtracted. Magnetic field lines are shown in the $\alpha_{M,r\phi}$ panel. The bottom panels show the quantities along the $\theta$ direction at r=$R_0$ and 2$R_0$, and negative values are shown as dotted curves. 
\label{fig:alpha}}
\end{figure*}  
  
It is computationally prohibitive to run a large suite of 3-D MHD simulations.  Thus, we would like to provide, or at least shed light on, stress and resistivity structures that enable 2-D axisymmetric simulations with anomalous viscosity and resistivity. The $r-\phi$ Reynolds and Maxwell stresses are shown in the left two columns of Figure \ref{fig:alpha}. It shows that $r^3\alpha_{R,r\phi}$ and $r^3\alpha_{M,r\phi}$ have different radial dependences, with the former increasing with radius and the latter decreasing with radius. At $r<2R_0$, the Maxwell stress is $\sim$2 orders of magnitude higher than the Reynolds stress. At the outer disk (e.g. r=3.5 $R_0$ midplane), the Maxwell stress is $\sim$1 order of magnitude higher, closer to the relationship in local MRI turbulent disks \citep{Hawley1995}. Laminar stress due to net magnetic fields dominates over the turbulent stress in the magnetically dominated region around the magnetospheric truncation radius. Only in the outer disk, turbulent stress becomes important. The Maxwell stress also becomes negative close to the current layer between the stellar and disk fields.

The $T_{\theta\phi}$ stress (in the fourth column of Figure \ref{fig:alpha}) affects the relative motion between different layers in the disk. The radial momentum of a region from [$\theta_{min}$, $\theta_{max}$] is contributed by $T_{\theta\phi}(\theta_{max})-T_{\theta\phi}(\theta_{min})$ (e.g. Equation 16 in \citealt{ZhuStone2018}). Thus, $T_{\theta\phi}$ leads to outflow in the midplane and inflow at the surface. This magnetic  breaking by the midplane to the surface is led by the radially tilted field lines resulting from the surface accretion.

Finally, to estimate the anomalous resistivity due to turbulence, we use the induction equation with the magnetic vector potential
\begin{equation}
\frac{\partial {\bold A}}{\partial t}={\bold v}\times{\bold B}-\eta_{turb}\nabla\times{\bold B}\,,
\end{equation}
where $\eta_{turb}$ is the resistivity due to turbulence. Since turbulent resistivity could be anisotropic, $\eta_{turb}$ should be a tensor. To simplify the analysis, we assume that $\eta_{turb}$ is isotropic. Then, for a steady state, we have
\begin{equation}
\eta_{turb}=\frac{{\bold v}\times{\bold B}}{\nabla\times{\bold B}}\,, \label{eq:eta}
\end{equation}
so that we can calculate $\eta_{turb}$ using three different components of ${\bold v}\times{\bold B}$ and $\nabla\times{\bold B}$.  $\eta_{turb}$ calculated using the $\phi$ component is shown in the rightmost panel of Figure \ref{fig:alpha}. In the disk region, this dimensionless $\eta$ value is between $\alpha_{R,r\phi}$ and $\alpha_{M,r\phi}$. On the other hand, we note that, if we use either $r$ or $\theta$ component in Equation \ref{eq:eta} to calculate the resistivity, we obtain much higher resistivitIes (more than 3 orders of magnitude higher). This is driven by the large $B_{\phi}$ value in Equation \ref{eq:eta}. This highly anisotropic $\eta$, together with the importance of $T_{\theta\phi}$, makes constructing accurate 2-D models challenging.   

\subsection{Comparison with Previous Works and Caveats}

Many results from previous 2-D simulations have been confirmed by our 3-D MHD simulations. For example, \cite{Zanni2013} revealed that field inflation could launch magnetospheric ejections that carry away angular momentum. \cite{Ustyugova2006} found powerful winds in the propeller regime, and \cite{Romanova2004b} detected accretion onto the star in this regime.  More quantitatively, for non-rotators and slow-rotators, our $n$ value approaches 1, which is consistent with analytical calculations \citep{Ghosh1979II} and 2-D simulations by \cite{Ireland2021}\footnote{$n=\dot{J}/(\dot{M}\sqrt{R_T/R_*})$ calculated with values in their Table 1.}.

The effects of stellar rotation on the interchange instability \citep{Blinova2016} are also observed in our simulations. The ``ordered unstable regime'' described by \cite{Blinova2016} appears similar to our non-rotator or the slowest rotator (Rc2) cases. The large-scale bubbles in Rc2 lead to one dominant tongue reaching the star. The periodogram also shows strong peaks at periods shorter than the stellar rotation periods. The ``chaotic unstable regime'' in \cite{Blinova2016} is similar to our Rc1 case, where multiple tongues are present, and the stellar period dominates in the periodogram. The $\omega_s\equiv(R_{T}/R_c)^{1.5}$ in our Rc2 and Rc1 cases (0.24, 0.49 respectively) are also consistent with the proposed separation of these two unstable regimes  $\omega_s\sim 0.45$  \citep{Blinova2016}. On the other hand, the separation between the unstable and stable regimes is $\omega_s\sim (0.85)^{1.5}=0.78$ in our simulations, slightly higher than 0.6 reported by \cite{Blinova2016}. Our equilibrium spin state with $\omega_s\sim 0.7$ are in the chaotic unstable regime with multiple tongues. Compared with previous works, the tongues/filaments in our simulations show finer structures (e.g., bifurcation), probably due to our higher resolution. We also observe that the large-scale magnetic bubbles likely link to the outflow, similar to accretion disks in the MAD state \citep{Tchekhovskoy2011}.

Our 3-D simulations show some significant differences from previous simulations.
First, our simulations reveal a thick magnetized disk that undergoes surface accretion. For fast rotators, surface material accretes from the outer disk to the inner disk, where it is channeled into the magnetosphere. Meanwhile, midplane material moves outward, originating from the magnetospheric truncation radius (Figure \ref{fig:torque2d}). For slow rotators, the midplane outflow begins at a radius larger than the truncation radius, so that the inner portion of the disk accretes inward across all heights. During surface accretion, the vertically differential motion drags magnetic fields into ``$\Sigma$'' shaped, which transports angular momentum from the surface to the midplane, reinforcing midplane outflow, similar to the MRI channel mode. This midplane outflow could potentially carry thermally processed materials (e.g. calcium-aluminium-rich inclusion, CAI) to the outer disk. 

 Second, due to frequent reconnection and the resulting turbulent diffusion, 
closed stellar field lines are transported well into the inner disk, leading to efficient star-disk coupling. For our fast rotators ($R_{T,ana}/R_c=2$), $n$ can be $\sim$ -10. The sensitive dependence of the spin-up/down torque on stellar rotation could suggest that most young stars should be around the equilibrium spin state. This is quite different from \cite{Matt2005} who argue that field inflation limits the star-disk coupling and stellar wind is necessary to explain observed stellar spin rates. On the other hand, we caution that most long-term disk evolutionary models to explain young star spins assume a smooth steady disk.  
The timescale of the stellar spin evolution can take thousands to millions of years. If the disk's accretion rate changes dramatically over a short period of time (e.g. like in our simulations),  the star could be in spin-up or spin-down state at any particular time.

 Third, both accretion and field inflation are less variable ($\sigma$ given in Table \ref{tab:summary}) compared with 2-D simulations. Field inflation may affect local accretion but cannot stop the accretion in all directions. Finally, compared with recent 3-D MHD simulations by \cite{Takasao2022} which show a highly distorted magnetosphere with one-sided magnetospheric accretion channels, our simulations show a highly magnetized corotating magnetosphere with two-sided accretion, similar to the traditional picture. While this discrepancy may be due to our smaller stellar radius compared with $R_T$, it may also be due to different stellar wind treatments, as discussed below.  

Our works has several caveats. First, we ignore the stellar wind. Magnetized stellar winds \citep{Cranmer2008,Cranmer2009} could extract angular momentum from the star \citep{Matt2005}. Since we have not adopted the realistic stellar temperature and density in the simulation, the density in our stellar wind region approaches the density floor, and the torque from this region is negligible. Second, thermodynamics has been simplified in our simulations. Radiation hydrodynamical simulations will be needed in future. However, simulating thermodynamics within the magnetosphere can be quite challenging since X-ray heating and wave heating need to be properly captured. Finally, we have not considered misalignment between the magnetic dipole and stellar rotation, and we have ignored external fields. Recent GRMHD simulations have highlighted the importance of these effects \citep{Parfrey2023,Murguia-Berthier2024}.

\subsection{Observational Implications}  
First, the derived equilibrium stellar spin matches the observed stellar rotation periods $\sim$1-10 days. Using the YSO parameters in \cite{Matt2005} with $\dot{M}=5\times 10^{-8}M_{\odot}/yr$, $M_{*}$=0.5$M_{\odot}$, $R_*=2R_{\odot}$, and $B_*=2kG$, the star in the equilibrium  spin state ($R_{T,ana}$=1.05 $R_c$) should have a period of 7.5 days. The equivalent breakup fraction $f\equiv\Omega_*\sqrt{R_*^3/GM_*}$ is 0.061, which is several times smaller than the spin of the equilibrium state in \cite{Matt2005}. Unlike \cite{Matt2005}, our simulation could already explain the observed YSO rotation without invoking stellar winds. 

Second, the covering factors of the hot spots in our simulations align with recent Hubble observations.  For example, \cite{Espaillat2022}  use both ULLYSES and  VLT X-Shooter data to  constrain that CVSO 109 has 5.5\%, 1.7\%, and 0.1\% covering factors for hot spots with energy fluxes of $10^{10}$ (low-energy), $10^{11}$ (mid-energy), and $10^{12}$ erg s$^{-1}$  cm$^{-2}$ (high-energy). To scale these hot spots to those in our simulations, we first calculate CVSO 109's mean energy flux $0.5 G M_*\dot{M}/(4\pi r_*^3)=3\times 10^9$ erg s$^{-1}$ cm$^{-2}$,  using the measured parameters of CVSO 109 in \cite{Espaillat2022} \footnote{$\dot{M}$ of 3.26$\times 10^{-8}M_{\odot}/yr$, $r_*$ of 1.8 $R_{\odot}$, and $M_*$ of 0.5 $M_{\odot}$}. Thus, CVSO 109's low-energy hot spots with $10^{10}$ erg s$^{-1}$ cm$^{-2}$ are roughly equivalent to hot spots in our simulations with 1 to 10 times the mean energy flux ($3\times 10^9$ to $3\times 10^{10}$  erg s$^{-1}$ cm$^{-2}$), shown as the green shaded region in Figure \ref{fig:sphere}. Similarly, mid-energy hot spots with $10^{11}$ erg s$^{-1}$ cm$^{-2}$ correspond to the yellow shaded region in the figure ($3\times 10^{10}$ to $3\times 10^{11}$  erg s$^{-1}$ cm$^{-2}$ for CVSO 109).  Depending on the spin rates, hot spots in the green region have covering factors of 5-20 $\%$, compared with 5.5$\%$ in the observation, while hot spots in the yellow region have covering factors of $\lesssim 5\%$, compared with 1.7$\%$ in the observations. Our simulated covering factors are also consistent with a bigger sample in \cite{Pittman2022, Wendeborn2024}. 

Third, the accretion variabilities are consistent with some observations. \cite{Herczeg2023} has collected the accretion rates of TW Hya over 25 years and found that the logarithm of its accretion rate has a Gaussian distribution with a standard deviation ($\sigma$) of 0.22 dex. This translates to $\sigma$ of $\dot{M}/\left<\dot{M}\right>$ $\sim$ 50\% in linear space. In our simulations, the ratio between the standard deviation and the mean accretion rate increases from 23$\%$ (1.02/4.37)\footnote{This ratio for the slow rotator is also consistent with the non-rotator limit of $\sim$ 23$\%$ in \cite{Zhu2024}.} to 52$\%$ (1.01/1.96) as the stellar spin increases (Table \ref{tab:summary}).   Thus, the relatively stable accretion in simulations matches observations (also for GM Aur in \citealt{Bouvier2023}).
  
Finallly, the episodic wind due to field inflation (Figure \ref{fig:mdottime}) seems to be consistent with recent GM Aur observations that its wind signatures randomly appear as blueshifted absorption components in the Balmer and HeI 10830 \AA  line profiles \citep{Bouvier2023}.  Our wind speed ($\sim 500$ km/s) and wind outflowing rate ($\sim10\%$ of $\dot{M}$ in the equilibrium spin state) are similar to the standard values in observations. More quantitative comparisons are needed in future.
  
Besides existing observations, our simulations make several predictions for future observations.

First, our simulations suggest that both the spatial and energy distributions of hot spots differ between slow and fast rotators. Slow rotators could have some hot spots around the equator due to the interchanging fingers/tongues, while fast rotators mainly have spots close to the pole (Figure \ref{fig:sphere}). Furthermore, the energy flux of hot spots is skewed towards higher values for faster rotators. As shown in the bottom panels of Figure \ref{fig:sphere}, the covering factor of higher energy flux hot spots (yellow) increases with faster stellar rotation.  Observationally, confirming these predictions requires a large sample of YSOs and may face some difficulties.
To determine if a star is a fast or slow rotator, we need to measure both the truncation radius and the corotation radius. While the corotation radius can be determined by studying the period of the star, the truncation radius requires careful line modeling \citep{Muzerolle1998}. 
 
Second, faster rotators' accretion should be more variable, and their wind rate/accretion rate ratio should be higher. Furthermore, the accretion may exhibit quasi-periodic signals due to the productions of magnetic bubbles and field inflation. Studying accretion and wind variability requires time-monitoring observations over tens to hundreds of stellar rotation periods. 
 
Third, for most stars, $R_{T}$ should be slightly smaller than $R_{c}$ ($\sim$0.79 $R_c$), if they are at the equilibrium spin state. However, during accretion outbursts (e.g. Exors and FUors), we may observe significant changes in $R_T$ before, during, and after the outbursts. The hot spot distribution and wind rate may also change accordingly in the outbursts. 

\section{Conclusion}
We extend our previous magnetospheric simulations by including rotating stars. After running for $\sim$ 500 Keplerian orbits at the stellar surface, our simulations have reached steady states for both the magnetosphere and the inner disk ($\lesssim$ 5 $R_T$).  The accretion is relatively steady, with a standard deviation ranging from 23$\%$ to 52$\%$ for our slow to fast rotators.
Similar to the non-rotator case, $\alpha$ values in the inner disks are high ($\alpha>$0.1 within 5 $R_T$), although  $\alpha$ and $\Sigma$ profiles become flatter around a moderately spinning star.

We first examined the accreting filaments within the magnetosphere. 
For slow rotators with $\omega_s\lesssim 0.78$ (equivalent to $R_{T}<0.85 R_c$ or  $R_{T,ana}<1.18 R_c$), the disk inner edge becomes unstable to interchange instability, allowing the intruding filaments/fingers to collide with the star near the equator, producing stellar hot spots around the equator. For fast rotators, interchange instability is suppressed, leading to a clean cavity inside the truncation radius, with most accretion following the dipole stellar magnetic fields and producing hot spots near the polar regions. The ``interchange instability'' criterion by \cite{Spruit1995} agrees with our simulations quite well, and the suppression of instability in fast rotators is primarily due to the super-Keplerian disk rotation outside the corotation radius. For slow rotators with instability, our simulations are consistent with the further division between the ``chaotic unstable regime'' and the ``ordered unstable regime'' \citep{Blinova2016}. Our simulation with $\omega_s=0.49$ shows multiple fingers within the magnetosphere, with the stellar period dominating in the periodogram, and thus is in the ``chaotic unstable regime''.
Our simulation with $\omega_s=0.24$ produces large-scale magnetic bubbles leading to one dominant finger and strong periodogram peaks at periods shorter than the stellar spin period, and is in the ``ordered unstable regime''.
 
At the stellar surface, infalling accretion columns have a wide range of  energy fluxes, creating hot spots with different temperatures. Most accretion energy is released from the low- and mid-energy flux hot spots. Low-energy flux hot spots are defined as those with infalling energy fluxes ranging from the mean energy flux value to 10 times higher, while mid-energy flux hot spots have fluxes 10-100 times the mean. The surface covering factor for low-energy flux hot spots is $\lesssim$ 20$\%$, while for mid-energy flux hot spots, it is $\lesssim$ 3 $\%$. For slow rotators, low-energy flux hot spots dominate, whereas fast rotators see higher contributions from mid-energy flux hot spots. In the Rc0p4 case, mid-energy flux hot spots cover $\sim 3\%$ of the surface but account for $\sim 60\%$ of the energy release.

After studying the magnetosphere, we analyze how the magnetospheric truncation radius varies with stellar spin.
For slow rotators, the truncation radius aligns well with $R_{T,ana}$ (as seen in \citealt{Zhu2024}), indicating that the ram pressure of accretion balances the magnetic pressure.  However, for fast rotators, the truncation radius is significantly smaller than $R_{T,ana}$ and closer to $R_{c}$. We provide a simple fitting formula for the truncation radius with respect to the stellar rotation. For fast rotators, some accreting material is directly channeled to the star, while some moves to the disk midplane and is centrifugally expelled out by the rotating stellar dipole. Overall, disks with $R_{T}$ significantly larger than $R_c$ should be rare, even for fast rotators in the so-called ``propeller'' regime.
 
Outside the truncation radius, our simulations reveal a thick magnetized disk undergoing surface accretion, which is different from most previous 2-D and 3-D simulations. At the disk surface, material can accrete smoothly into the magnetosphere channels. Since surface accretion drags magnetic fields inward,
the magnetic fields in the disk are ``$\Sigma$'' shaped. This field configuration transports angular momentum from the surface to the midplane, driving midplane outflow at large radii for all our cases. At the disk surface around $z\sim R$, field inflation generates episodic winds. Disk wind is $\sim$20$^o-$50$^o$ from the pole, while the stellar wind is within 20$^o$ from the pole.  This current-layer driven disk wind originates near the  magnetospheric truncation radius, and the wind mass loss rate increases from $\sim1\%$ to $\sim 40\%$ of the accretion rate as the star spins up.

Due to frequent reconnection and the resulting turbulent diffusion, 
closed stellar field lines extend well into the inner disk, leading to efficient star-disk coupling. We fit the stellar spin-up/down torque with respect to $R_T/R_{c}$, with the $n$ parameter ranging from 1 to -10 in our simulations. 
The magnetic field topology can be roughly divided into three regions: one with closed field lines around the star, one with open field lines anchored on the star, and one with open field lines threading through the disk. Since there are only dipole stellar fields initially, the open field lines are opened by field inflation at $z\sim R$. Following the open field lines to the disk midplane, they tilt outwards. Turbulent diffusion can transport closed stellar field lines to  $R\sim 2 - 3 R_0$, well beyond the truncation radius. Through these field lines, the stellar rotation drives the midplane outflow which could potentially carry thermally processed materials (e.g. CAI) to the outer disk.  

During the equilibrium spin state, $R_{c}\sim R_{T,ana} \sim 1.27 R_{T}$ or $\omega_s\sim 0.7$, suggesting a typical YSO in this state has a $\sim$7 day period even without any angular momentum transport by stellar winds. In this state, filaments are present within the magnetosphere, winds can reach speeds of up to 500 $km/s$, and the ratio between the wind outflowing rate and disk accretion rate is $\sim 13\%$. 

Our results on the surface accretion, the interchange instability criterion, outflow-$R_T$ relation, and torque-$R_T$ relation also stand for a thinner disk. 

Our simulations align with several existing observations, including YSOs' rotation periods, hot spot covering factors, accretion variabilities, episodic winds, wind speeds, and wind mass loss rates. Finally, we make testable predictions for future observations, particularly how these observables should change with the stellar spin rate.

\section*{DATA AVAILABILITY}
The data underlying this article will be shared on reasonable request to the corresponding author.

\acknowledgments
All  simulations are carried out using with NASA Pleiades supercomputer. Z. Z. acknowledges support from NASA award 80NSSC22K1413. ZZ thanks Caeley Pittman, Catherine Espaillat, Thanawuth Thanathibodee, and John Wendeborn for discussions. ZZ thanks Ian Rabago for his assistance in rendering Figure 16 using Blender.

\begin{figure*}[t!]
\centering
\includegraphics[trim=0mm 10mm 0mm 0mm, clip, width=5.in]{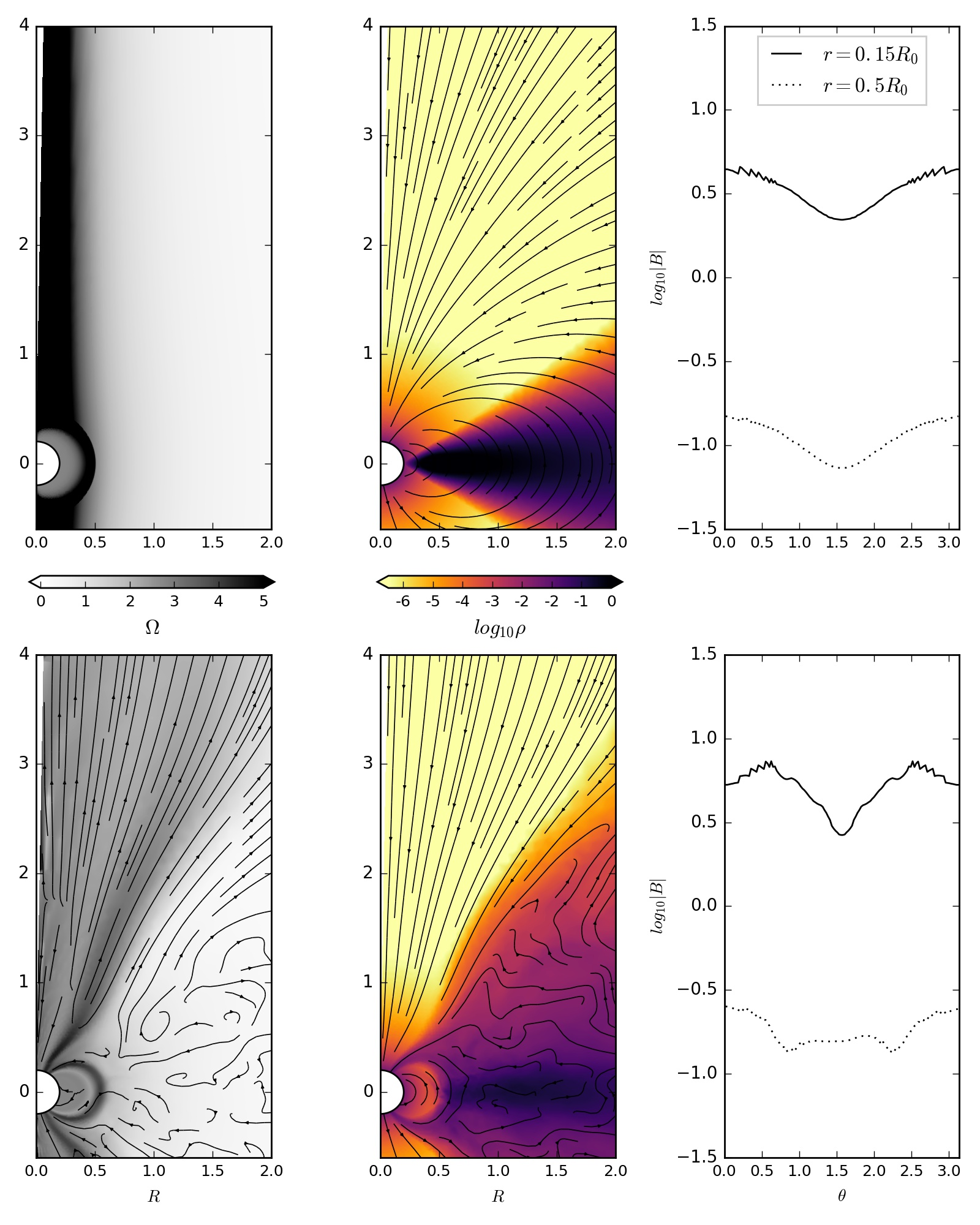}
\figcaption{The left and middle columns: the angular frequency $\Omega$ and density in the initial condition (upper panels) and the final state (bottom panels) of the $R_c=0.5 R_0$ case.  Quantities are azimuthally averaged.  The streamlines of the poloidal velocity are shown in the leftmost column (zero poloidal velocity in the initial condition), while the streamlines of the poloidal magnetic field are shown in the middle column. The right column: the magnetic field strength with respect to $\theta$ at r=0.15$R_0$ and 0.5$R_0$.
\label{fig:refereeini}}
\end{figure*}

\begin{figure*}[t!]
\centering
\includegraphics[trim=0mm 10mm 0mm 0mm, clip, width=5.in]{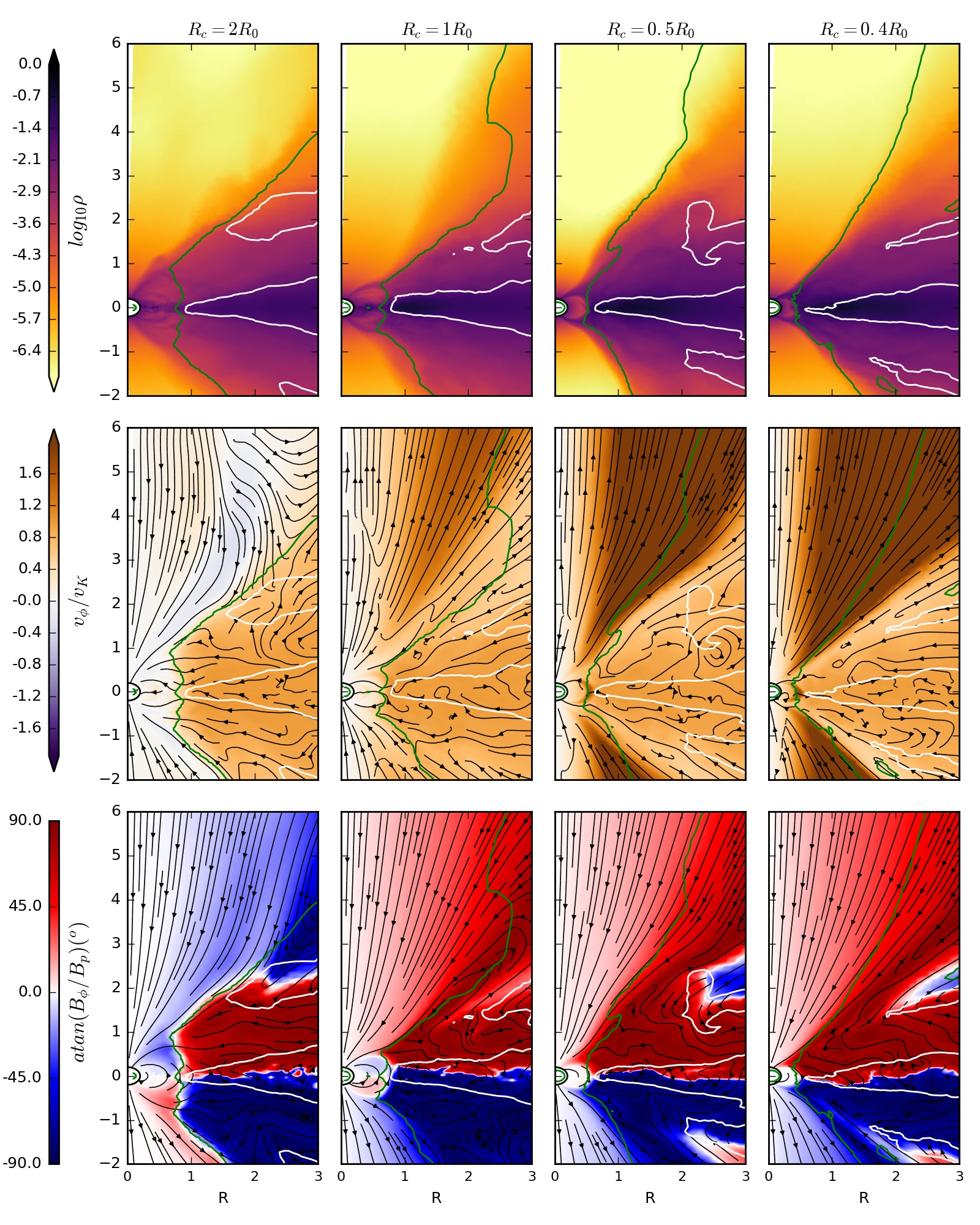}
\figcaption{ 
Density (top row), velocity (middle row), and magnetic fields (bottom row)  at the end of the simulations for all four cases. The streamlines in the middle and bottom rows represent the poloidal velocity and magnetic fields, respectively. White contours indicate where $\beta$=1, and the green curves mark where $E_k=E_B$. All primitive variables are azimuthally averaged.  
\label{fig:bubble}}
\end{figure*}

\appendix
The disk structure in the initial condition and at the end of the simulation for the $R_c=0.5 R_0$ case are shown in Figure \ref{fig:refereeini}.
The structures of density, velocity, and magnetic fields for all the cases are shown in Figure \ref{fig:bubble}.

\end{document}